\documentclass[12pt,cite,a4paper]{article}

\usepackage{amsmath}
\usepackage{amssymb}
\usepackage[dvips]{graphicx}
\numberwithin{equation}{section}
\renewcommand{\thefootnote}{\fnsymbol{footnote}}

\begin{document}
\begin{titlepage}
\begin{flushright}
IFUP--TH/2006\,--\,6\\
\end{flushright}
~

\vskip .8truecm
\begin{center}
\Large\bf
Liouville field theory with heavy charges.\\
I. The pseudosphere\footnote{This work is  supported in part
  by M.I.U.R.}
\end{center}

\vskip 1.6truecm
\begin{center}
{Pietro Menotti} \\
{\small\it Dipartimento di Fisica dell'Universit{\`a}, Pisa 56100,
Italy and}\\
{\small\it INFN, Sezione di Pisa}\\
{\small\it e-mail: menotti@df.unipi.it}\\
\end{center}
\vskip .8truecm
\begin{center}
{Erik Tonni} \\
{\small\it Scuola Normale Superiore, Pisa 56100, Italy and}\\
{\small\it INFN, Sezione di Pisa}\\
{\small\it e-mail: e.tonni@sns.it}\\
\end{center}

\vskip 1.6truecm

\begin{abstract}

We work out the perturbative expansion of quantum Liouville theory
on the pseudosphere starting from the semiclassical limit of a
background generated by heavy charges.  By solving perturbatively
the Riemann-Hilbert problem for the Poincar\'e accessory
parameters, we give in closed form the exact Green function on the
background generated by one finite charge.  Such a Green function is
used to compute the quantum determinants i.e. the one loop
corrections to known semiclassical limits thus providing the
resummation of infinite classes of standard perturbative graphs.
The results obtained for the one point function are compared with
the bootstrap formula while those for the two point function are
compared with the existing double perturbative expansion and with
a degenerate case, finding complete agreement.

\end{abstract}

\vskip 1truecm

\renewcommand{\thefootnote}{\arabic{footnote}}
\end{titlepage}

\section*{Introduction}

The conformal bootstrap program has provided very deep results in
conformal field theory and in particular in Liouville quantum
field theory \cite{ZZpseudosphere,FZZ,ZZsphere}. In this approach
one assumes at the outset conformal invariance and, by using
formal properties of the functional integral and some other
assumptions, one arrives at functional equations for the
correlation functions. Under reasonable regularity assumptions
their solution provide the exact correlation functions.\\
Here and in an accompanying paper we address the problem to
recover the conformal theory from the usual field theoretic
procedure in which one starts from a stable background and then
one integrates over the quantum fluctuations. As it is well known,
a quantum field theory is specified not only by an action but also
by a regularization and renormalization procedure. In
\cite{MTgeometric,MV,DFJ} it was found that not all the
regularization procedures provide a theory which is invariant
under the full conformal group. The regularization suggested at
the perturbative level in \cite{ZZpseudosphere} in the case of the
pseudosphere provides the vertex functions with the correct
quantum dimensions \cite{CT} at the first perturbative order. Here
it is explicitly proved that such a result stays unchanged to all
orders perturbation theory. In particular the weight of the
cosmological term becomes $(1,1)$ as required by the invariance
under local conformal transformations.\\
The pseudosphere case was already considered in
\cite{ZZpseudosphere} and more fully developed in
\cite{MTgeometric,MTtetrahedron}. These calculations correspond to
a double perturbative expansion in the coupling constant and in
the charge of the vertex function. \\
Here instead we start from the background generated by finite
charges, i.e. ``heavy charges'' in the terminology of
\cite{ZZsphere}. This means that we consider the vertex operators
$V_{\alpha_n}(z_n) = e^{2\alpha_n \phi(z_n)}$
with $\alpha_n=\eta_n/b$ and $\eta_n$ fixed in the
semiclassical limit $b \rightarrow 0$.\\
In the case of a single heavy charge, by solving a Riemann-Hilbert
problem we are able to compute the exact Green function on such a
background in closed form, and the Green function is used to
develop the subsequent perturbative expansion in the coupling
constant for the one and two point functions.
In this way we obtain a resummation of infinite classes
of perturbative graphs.

\noindent Some of the results derived here were reported in
\cite{MTheavyZZ}. In the present paper we give full details of the
computational procedure. In section \ref{classical} we lay down
the notations and discuss the semiclassical limit. In section
\ref{quantum dimensions sec} it is shown that the regularization
procedure of \cite{ZZpseudosphere} provides the vertex functions
with the correct quantum dimensions to all orders perturbation
theory. In section \ref{Greenfunctionsect} we solve the Riemann
Hilbert problem which allows the determination of the exact Green
function on the background given by one heavy charge. In section
\ref{quantum determinant sec} the one loop correction to the
semiclassical one point function is computed. The result is
compared with the expansion of the exact one point function
derived in the bootstrap approach \cite{ZZpseudosphere} finding
complete agreement. In section \ref{two point sec} the two point
function with one finite charge and an infinitesimal one is
computed by employing analogous technique. Particular cases of
such new result are compared with the existing double perturbative
expansion in $\alpha$ and $b$ and with a degenerate case, finding
complete agreement in both cases. In appendix \ref{appendix phiB}
we derive the behavior of the conformal factor for the $N$ point
classical background at infinity and in appendix \ref{details
green function} we give the details of the computation of the
Green function.

\section{Classical and quantum action on the pseudosphere}
\label{classical}

Let us consider the geometry of the pseudosphere in the
representation of the unit disk $\Delta = \{z \in\mathbb{C}\,; |z|
< 1\}$.\\
We write the $N$ point function for the vertex operators
$V_\alpha(z) = e^{2\alpha \phi(z)}$ in the form
\begin{equation}\label{N point geometric}
\left\langle \, e^{2\alpha_1 \phi(z_1)} \dots\, e^{2\alpha_N
\phi(z_N)}\,\right\rangle
 \,=\,\frac{1}{Z}\;\int\hspace{-.06cm}
\mathcal{D}\,[\, \phi \,]\;\, e^{\,-S_{\Delta,N}[\,\phi\,]}
\end{equation}
where
\begin{equation}\label{partition function}
Z
 \,=\,\int\hspace{-.06cm}
\mathcal{D}\,[\, \phi \,]\;\, e^{\,-S_{\Delta,N=\,0}[\,\phi\,]}
\end{equation}
and $S_{\Delta,\,N}[ \,\phi\,]$ is the action of Liouville field
theory on the pseudosphere in presence of $N$ sources, which is
given by the following expression \cite{ZZsphere,MTgeometric}
\begin{eqnarray}\label{geometric action}
\hspace{-1.7cm} S_{\Delta,\,N}[ \,\phi\,] & = &
 \lim_{\begin{array}{l}
\vspace{-.9cm}~\\
\hspace{0cm} \vspace{-.4cm} \scriptscriptstyle \varepsilon_n \rightarrow\, 0 \\
\hspace{.2cm} \scriptscriptstyle \!\!r\;
\rightarrow\,1\end{array}}\, \Bigg\{
\int_{\Delta_{\,r,\,\varepsilon}} \left[ \,\frac{1}{\pi}
\,\partial_z \phi \,\partial_{\bar{z}}\phi+\mu\,
e^{2b\phi}\,\right]\, d ^2 z \,  \nonumber \\
  &  &\hspace{1.4cm} -\,\frac{Q}{2\pi i}\; \oint_{\partial\Delta_r} \phi
\left( \, \frac{\bar{z}}{1-z\bar{z}} \,d z-
       \,\frac{z}{1-z\bar{z}} \,d \bar{z} \right)+ f(r,b)\nonumber\\
  &  & \hspace{1.4cm}\rule{0pt}{1cm}
   -\,\frac{1}{2\pi i}\;\sum_{n=1} ^N \alpha_n\oint_{\partial\gamma_n} \phi
\left( \, \frac{d z}{z-z_n}- \frac{d \bar{z}}{\bar{z}-\bar{z}_n}\,
\right)- \sum_{n=1} ^N \alpha_n^2 \log \varepsilon_n^2 \; \Bigg\}
\end{eqnarray}
with $Q=1/b + b$ and $d^2 z=idz\wedge d\bar{z}/2$.\\
The integration domain
$\Delta_{\,r,\,\varepsilon}=\Delta_{\,r}\hspace{-.07cm} \setminus
\bigcup_{n=1}^N \gamma_{n}$ is obtained by removing $N$ disks
$\gamma_{n}=\{|z-z_n|<\varepsilon_n\}$ from the disk
$\Delta_{\,r}=\{\,|z| < r < 1 \,\} \subset \Delta$.
The boundary behaviors of the field $\phi$ are given by
\begin{eqnarray}
\label{phi bc infinity} \phi(z) & = & -\,\frac{Q}{2}\, \log\,
(\,1-z\bar{z}\,)^2 + O(1)
\hspace{.7cm}\textrm{when}\hspace{.7cm} |z|\rightarrow 1 \\
\label{phi bc sources} \phi(z)  & = & -\,\alpha_n \, \log
\,|\,z-z_n\,|^2\,+O(1)
\hspace{.7cm}\textrm{when}\hspace{.7cm}\,z\,\rightarrow z_n\;.
\end{eqnarray}
The function $f(r,b)$ is a subtraction term independent of
the field $\phi$ and of the charges.\\

\noindent In order to connect the quantum theory to its
semiclassical limit it is useful to define \cite{ZZsphere}
\begin{equation}
\varphi \,=\, 2 b \phi\;, \hspace{1.5cm}
\alpha_n\,=\,\frac{\eta_n}{b}\;.
\end{equation}
The charges $\alpha_n = \eta_n/b$ are called heavy charges
\cite{ZZsphere} because in the perturbative limit $b\rightarrow 0$
 the parameters $\eta_n$ are kept fixed and therefore $\alpha_n$
 diverge. Since the measure is $e^{\varphi} d^2z$, the condition of local
finiteness of the area around each source and the asymptotic
behavior
(\ref{phi bc sources}) for the field $\phi$ impose that
$1-2 \eta_n>0$ \cite{Seiberg: Notes,Picard}.\\
Now we decompose the field $\varphi$ as the sum of a classical
background field $\varphi_{\scriptscriptstyle\hspace{-.05cm}B}$
and a quantum field
\begin{equation}
\varphi \,=\,\varphi_{\scriptscriptstyle\hspace{-.05cm}B} +
2b\,\chi\;.
\end{equation}
Then, we can write the action as the sum of a classical and a
quantum action as follows
\begin{equation}\label{classicalaction}
S_{\Delta,\,N}[ \,\phi\,]  \,=\,S_{cl}[
\,\varphi_{\scriptscriptstyle\hspace{-.05cm}B}\,] +
S_{q}[\,\varphi_{\scriptscriptstyle\hspace{-.05cm}B},\,\chi\,]\;.
\end{equation}
The classical action is given by
\begin{eqnarray}\label{action classical phiB}
\hspace{0cm} S_{cl}[
\,\varphi_{\scriptscriptstyle\hspace{-.05cm}B}\,] & =
&\frac{1}{b^2}
 \lim_{\begin{array}{l}
\vspace{-.9cm}~\\
\hspace{0cm} \vspace{-.4cm} \scriptscriptstyle \varepsilon_n \rightarrow\, 0 \\
\hspace{.2cm} \scriptscriptstyle \!\!r\;
\rightarrow\,1\end{array}}\,
\Bigg\{\int_{\Delta_{r,\varepsilon}}
\left[ \,\frac{1}{4\pi} \,\partial_z
\varphi_{\scriptscriptstyle\hspace{-.05cm}B}
\,\partial_{\bar{z}}\varphi_{\scriptscriptstyle\hspace{-.05cm}B}+\mu
b^2 e^{\,\varphi_{\hspace{-.02cm}
\scriptscriptstyle{B}}}\,\right]\, d^2 z \\
  &  &\hspace{2.2cm} -\,\frac{1}{4\pi i}\; \oint_{\partial\Delta_r}
  \varphi_{\scriptscriptstyle\hspace{-.05cm}B}
\left( \, \frac{\bar{z}}{1-z\bar{z}} \,d z-
       \,\frac{z}{1-z\bar{z}} \,d \bar{z} \right)+ f_{cl}(r,\mu b^2)\nonumber\\
 \rule{0pt}{1cm} &  & \hspace{2.2cm}
   -\,\frac{1}{4\pi i}\;\sum_{n=1} ^N \eta_n\oint_{\partial\gamma_n}
   \varphi_{\scriptscriptstyle\hspace{-.05cm}B}
\left( \, \frac{d z}{z-z_n}- \frac{d \bar{z}}{\bar{z}-\bar{z}_n}\,
\right)- \sum_{n=1} ^N \eta_n^2 \log \varepsilon_n^2 \; \Bigg\}
\nonumber
\end{eqnarray}
while the quantum action reads
\begin{eqnarray}\label{quantumaction}
\hspace{0cm} S_{q}[
\,\varphi_{\scriptscriptstyle\hspace{-.05cm}B}\,, \chi\,] & = &
 \lim_{\begin{array}{l}
\vspace{-.9cm}~\\
\hspace{0cm} \vspace{-.4cm} \scriptscriptstyle \varepsilon_n \rightarrow\, 0 \\
\hspace{.2cm} \scriptscriptstyle \!\!r\;
\rightarrow\,1\end{array}}\, \Bigg\{ \int_{\Delta_{r,\varepsilon}}
\left[ \,\frac{1}{\pi} \,\partial_z \chi
\,\partial_{\bar{z}}\chi+\mu\, e^{\,\varphi_{\hspace{-.02cm}
\scriptscriptstyle{B}}}\,\big(\,e^{2b\,\chi}-1\,\big)
-\frac{1}{\pi
b}\;\chi\,\partial_z\partial_{\bar{z}}\varphi_{\scriptscriptstyle
\hspace{-.05cm}B}\,\right]\, d^2 z
 \nonumber \\
& & \hspace{-1.6cm}-\,\frac{1}{2\pi i\,b}\;
\oint_{\partial\Delta_r}
  \chi\,
\left( \, \frac{\bar{z}}{1-z\bar{z}} \,-\, \frac{1}{2}\,\partial_z
\varphi_{\scriptscriptstyle\hspace{-.05cm}B}\right)\,dz
\,+\,\frac{1}{2\pi i\,b}\; \oint_{\partial\Delta_r}
  \chi\,
\left( \, \frac{z}{1-z\bar{z}} \,-\,
\frac{1}{2}\,\partial_{\bar{z}}
\varphi_{\scriptscriptstyle\hspace{-.05cm}B}\right)\,d\bar{z}
 \nonumber \\
 \rule{0pt}{1cm}   &  &
\hspace{1.4cm}
  -\,\frac{1}{4\pi i}\; \oint_{\partial\Delta_r}
  \varphi_{\scriptscriptstyle\hspace{-.05cm}B}\,
\left( \, \frac{\bar{z}}{1-z\bar{z}} \,d z\,- \frac{z}{1-z\bar{z}}
\,d \bar{z} \right)+ f_{q}(r,\mu b^2)
       \nonumber \\
\rule{0pt}{1cm}  &  &\hspace{1.4cm} -\,\frac{b}{2\pi i}\;
\oint_{\partial\Delta_r}
  \chi\,
\left( \, \frac{\bar{z}}{1-z\bar{z}} \,d z\,- \frac{z}{1-z\bar{z}}
\,d \bar{z} \right)\\
\rule{0pt}{1cm} & & \hspace{-2.8cm} -\,\frac{1}{2\pi i\,b}\,
\sum_{n=1} ^N \;\oint_{\partial\gamma_n} \hspace{-.1cm} \chi
\left( \, \frac{\eta_n}{z-z_n} \,+\, \frac{1}{2}\,\partial_z
\varphi_{\scriptscriptstyle\hspace{-.05cm}B}\right)dz
\,+\,\frac{1}{2\pi i\,b}\, \sum_{n=1} ^N\;
\oint_{\partial\gamma_n}\hspace{-.1cm} \chi \left( \,
\frac{\eta_n}{\bar{z}-\bar{z}_n}\,+\,
\frac{1}{2}\,\partial_{\bar{z}}
\varphi_{\scriptscriptstyle\hspace{-.05cm}B}\right)d\bar{z}
       \;\Bigg\}\,. \nonumber
\end{eqnarray}
We remark that
the subtraction terms $f_{cl}(r,\mu b^2)$ and $f_{q}(r,\mu b^2)$
are independent
of the fields and of the charges $\eta_n$.\\
For the classical background field near the sources we have
\begin{equation}\label{varphiB bc sources}
\varphi_{\scriptscriptstyle\hspace{-.05cm}B}(z) \, =\, -\;2\eta_n
\log |z-z_n|^2\,+O(1) \hspace{1cm} \textrm{when}
\hspace{.44cm}\,z\,\rightarrow z_n
\end{equation}
while in appendix \ref{appendix phiB} the following boundary
behavior for $\varphi_{\scriptscriptstyle\hspace{-.05cm}B}(z)$ is
proved
\begin{equation}
\label{varphiB bc infinity}
\varphi_{\scriptscriptstyle\hspace{-.05cm}B}(z)\, =\, -\, \log\,
(1-z\bar{z})^2 + f(\mu b^2) + O\big((1-z\bar{z})^2\big)
\hspace{.9cm} \textrm{when} \hspace{.4cm} |z|\rightarrow 1
\end{equation}
where $f(\mu b^2)$ is a constant depending on $\mu b^2$.\\
Comparing (\ref{varphiB bc sources}) with (\ref{phi bc sources}),
we see that $\chi$ is regular at the sources. This fact and the
boundary behavior (\ref{varphiB bc sources}) imply the vanishing
of the last line in (\ref{quantumaction}) in the limit
$\varepsilon_n \rightarrow 0$. Moreover, since the field $\chi$
can diverge only like a logarithm when $z\bar{z} \rightarrow 1$,
the asymptotic (\ref{varphiB bc infinity}) implies that the second
line in (\ref{quantumaction}) vanishes in the limit $r \rightarrow
1$.

\noindent Now we focus on the classical action
$S_{cl}[\,\varphi_{\scriptscriptstyle\hspace{-.05cm}B}\,]$. The
vanishing of its first variation  with respect to the field
$\varphi_{\scriptscriptstyle\hspace{-.05cm}B}$ with boundary
conditions (\ref{varphiB bc infinity}) and (\ref{varphiB bc
sources}) gives the Liouville equation in presence of $N$ sources
\begin{equation}\label{liouville eq with sources varphi}
    -\,\partial_{z}\partial_{\bar{z}}\,
\varphi_{\scriptscriptstyle\hspace{-.05cm}B}\,
    +
    2\pi\,\mu b^2\,e^{\,\varphi_{\hspace{-.02cm}
    \scriptscriptstyle{B}}} \,=\,
    2\pi \sum_{n=1}^N \eta_n
    \,\delta^2(z-z_n)\;.
\end{equation}
Under a generic conformal transformation $z \rightarrow w(z)$ the
background field changes as follows
\begin{equation}\label{varphiB transformations}
\varphi_{\scriptscriptstyle\hspace{-.05cm}B}(z)
\hspace{.5cm}\longrightarrow  \hspace{.5cm}
\tilde{\varphi}_{\scriptscriptstyle\hspace{-.05cm}B}(w)
\;=\;\varphi_{\scriptscriptstyle\hspace{-.05cm}B}(z)\,-\, \log
\left| \frac{d w}{d z}\right|^{2}
\end{equation}
so that $e^{\,\varphi_{\hspace{-.02cm}
    \scriptscriptstyle{B}}} d^2z$ is invariant.\\
In particular, under a $SU(1,1)$ transformation, which maps the
unit disk $\Delta$ into itself, the classical action (\ref{action
classical phiB}) changes as follows \cite{MTgeometric,Takhtajan}
\begin{equation}\label{SU(1,1) transf law classic}
\widetilde{S}_{cl}[
\,\tilde{\varphi}_{\scriptscriptstyle\hspace{-.05cm}B}\,]\,=\,
S_{cl}[ \,\varphi_{\scriptscriptstyle\hspace{-.05cm}B}\,] +
\sum_{n\,=\,1}^N  \frac{\eta_n (\,1-\eta_n)}{b^2}\; \log\left|
\frac{d w}{d z}\right|_{z\,=\,z_n}^2\;.
\end{equation}
The classical action $S_{cl}[
\,\varphi_{\scriptscriptstyle\hspace{-.05cm}B}\,]$ computed on the
solution $\varphi_{\scriptscriptstyle\hspace{-.05cm}B}$ of the
equation of motion (\ref{liouville eq with sources varphi})
becomes a function $S_{cl}(\eta_1,z_1;\dots;\eta_N,z_N)$ of the
positions $z_n$ of the sources and of their charges $\eta_n$.
This function provides the semiclassical expression
of the $N$ point
function for the Liouville vertex operators $V_\alpha(z) =
e^{2\alpha\phi(z)}$
\begin{equation}\label{semiclassic N point}
\left\langle \, V_{\alpha_1}(z_1)\dots V_{\alpha_N}(z_N)
\,\right\rangle_{sc}\,=\,
\frac{e^{-S_{cl}(\eta_1,\,z_1;\,\dots\,;\,\eta_N,\,z_N)}}{e^{-S_{cl}(0)}}\;.
\end{equation}
The function in the denominator is the semiclassical contribution
of the partition function $Z$ in (\ref{N point
geometric}) and it is $SU(1,1)$ invariant.\\
By using (\ref{SU(1,1) transf law classic}), we immediately see
that (\ref{semiclassic N point}) has the following transformation
properties under $SU(1,1)$
\begin{equation}\label{semiclassic trasf N point}
\langle \, \widetilde{V}_{\alpha_1}(w_1)\dots
\widetilde{V}_{\alpha_N}(w_N) \,\rangle_{sc}\,=\,
\prod_{n\,=\,1}^N \left|\frac{d w}{dz} \right|_{z=z_n}^{- \,2\,
\eta_n (1-\eta_n)/b^2} \hspace{-.2cm}\langle \,
V_{\alpha_1}(z_1)\dots V_{\alpha_N}(z_N) \,\rangle_{sc}\;.
\end{equation}
This means that the semiclassical dimensions of the vertex
operator $V_\alpha(z)$ are
$ \eta(1- \eta)/b^2 = \alpha\,(1/b - \alpha)$.\\
Now we consider the quantum action (\ref{quantumaction}). For a
background field $\varphi_{\scriptscriptstyle\hspace{-.05cm}B}$
satisfying the Liouville equation with sources (\ref{liouville eq
with sources varphi}) and the boundary conditions (\ref{varphiB bc
infinity}) and (\ref{varphiB bc sources}), the quantum action
(\ref{quantumaction}) becomes
\begin{eqnarray}\label{quantum action final}
\vspace{-1cm}
 S_{q}[\,\varphi_{\scriptscriptstyle\hspace{-.05cm}B}\,, \chi\,] & = &
 \lim_{\scriptstyle r\rightarrow 1}
\, \Bigg\{ \int_{\Delta_{r}} \left[ \,\frac{1}{\pi} \,\partial_z
\chi \,\partial_{\bar{z}}\chi+\mu\, e^{\,\varphi_{\hspace{-.02cm}
\scriptscriptstyle{B}}}\,\big(\,e^{2b\,\chi}-1-2b\,\chi\,\big)\,\right]\,
d^2 z\\
 &  &\hspace{4.8cm} -\,\frac{b}{2\pi i}\;
\oint_{\partial\Delta_r}
  \chi\,
\left( \, \frac{\bar{z}}{1-z\bar{z}} \,d z\,- \frac{z}{1-z\bar{z}}
\,d \bar{z} \right)
       \;\Bigg\}\;. \nonumber
\end{eqnarray}
If the Green function vanishes quadratically on the boundary, the
last term in (\ref{quantum action
  final}) does not contribute to the perturbative expansion. In section
  \ref{Greenfunctionsect} we shall verify this fact
  explicitly for the case $N=1$.\\

\noindent In this paper, we are mainly interested in the case of a
single source, i.e. $N=1$ and $\eta_1=\eta$. By exploiting the
invariance under $SU(1,1)$ we can place the source in $z_1=0$. In
this case the background field, i.e. the solution of the Liouville
equation (\ref{liouville eq with sources varphi}) with boundary
behaviors (\ref{varphiB bc infinity}) and (\ref{varphiB bc
sources}), can be explicitly written \cite{Seiberg: Notes,
GinspargMoore}
\begin{equation}\label{phiclassic}
    e^{\varphi_{cl}}\,=\,\frac{1}{\pi\mu b^2}\;
    \frac{(1-2\eta)^2}{\big(\rule{0pt}{.4cm}
(z\bar{z})^{\eta}-(z\bar{z})^{1-\eta}\big)^2}\;\,.
\end{equation}
It is important to notice that the behavior of $\varphi_{cl}$ on
the boundary $\partial \Delta$, i.e. at infinity, is independent
of $\eta$ both in the divergent term and in the constant term
\begin{equation}\label{b.c. phicl infinity}
\varphi_{cl} \,=\, -\, \log(1-z\bar{z})^2 -\,\log\big(\pi \mu
b^2\big)\,+ \,O\big((1-z\bar{z})^2\big)
\hspace{.8cm}\textrm{when}\hspace{.7cm} z\bar{z} \rightarrow 1\;.
\end{equation}
Notice that the term $O(1-z\bar{z})$ is also absent, in agreement
with the asymptotics (\ref{varphiB bc infinity}) for
the background field.\\
In appendix \ref{appendix phiB} we prove that this is a general
feature for the solution
$\varphi_{\scriptscriptstyle\hspace{-.05cm}B}$ in presence of $N$
sources; therefore the two boundary integrals in the second line
of (\ref{quantumaction}) vanish when $|z|\rightarrow 1$,
being $\chi$ logarithmically divergent at most.\\
\noindent By using the explicit form of the classical background
field (\ref{phiclassic}), we can write the expression of the
semiclassical one point function for a vertex operator with charge
$\alpha=\eta/b$ placed in $0$
\begin{eqnarray}\label{one point classical term}
\left\langle \, V_{\eta/b}(0) \,\right\rangle_{sc} & = &
\exp\left\{-\,S_{cl}[ \,\varphi_{cl}\,]\,+\,\left.S_{cl}[
\,\varphi_{cl}\,]\right|_{\,\eta\,=\,0}\,\right\}\\
\rule{0pt}{.8cm}& = &
\exp\left\{-\,\frac{1}{b^2}\;\Big(\,\eta\,\log\big(\pi \mu
b^2\big)+2\eta+(1-2\eta)\,\log(1-2\eta)\,\Big)\,\right\}\;.
        \nonumber
\end{eqnarray}
The one point function in the basic vacuum found by Zamoldchikov
and Zamolodchikov \cite{ZZpseudosphere} within the conformal
bootstrap approach is
\begin{equation}
 \label{one-point function}
\left\langle \,V_{\alpha}(z_1) \,\right\rangle =
\frac{U_{1,1}(\alpha)}{(1-z_1\bar{z}_1)^{\,2\alpha(Q-\alpha)}}
\end{equation}
with
\begin{equation}\label{bootstrap ZZ onepoint}
U_{1,1}(\alpha)\,=\,\big(\pi\mu\gamma(b^2)\big)^{-\alpha/b}
\frac{\Gamma(Q\,b)\,\Gamma(Q/b)\,Q}{\rule{0pt}{.36cm}
  \Gamma\big((Q-2\alpha)\,b\big)\,\Gamma\big((Q-2\alpha)/b\big)\,(Q-2\alpha)} 
\end{equation}
where $Q=1/b+b$ and $\gamma(x)=\Gamma(x)/\Gamma(1-x)$.\\
The expression (\ref{one point classical term}) agrees with the
semiclassical term of (\ref{one-point function}) for $z_1=0$ and
$\alpha=\eta/b$.

\section{The quantum dimensions}
\label{quantum dimensions sec}

\noindent In this section we show that the quantum determinant of
the $N$ point function provides the quantum correction to the
conformal dimensions and that no further contributions to the
conformal dimensions occur.\\
The $O(b^0)$ quantum correction to the $N$ point function
$\left\langle \, V_{\alpha_1}(z_1)\dots V_{\alpha_N}(z_N)
\,\right\rangle$ is given by the quantum determinant
\begin{equation}\label{det definition}
\big(\,\textrm{Det}\,D\,\big)^{-1/2}\,\equiv\,\frac{1}{Z_0}\int\hspace{-.06cm}
\mathcal{D}\,[\, \chi \,]\;\,
\exp\left\{-\,\frac{1}{2}\,\int_\Delta \chi
\left(-\,\frac{2}{\pi}\,\partial_z\partial_{\bar{z}}\,+ \,4\mu
    b^2\,e^{\varphi_{\scriptscriptstyle\hspace{-.05cm}B}}
\right)\chi\,d^2z\,\right\}
\end{equation}
where $\varphi_{\scriptscriptstyle\hspace{-.05cm}B}$ is the
classical background field solving the Liouville equation and with
asymptotics (\ref{varphiB bc infinity}) and (\ref{varphiB bc
sources}), while $Z_0$ is the quadratic part of the
partition function (\ref{partition function}).\\
Taking the logarithmic derivative w.r.t. $\eta_j$ of
$\big(\,\textrm{Det}\,D\,\big)^{-1/2}$, we find the following
integral
\begin{equation}\label{det integral}
\frac{\partial}{\partial \eta_j}\,\log
\big(\,\textrm{Det}\,D\,\big)^{-1/2} =\;-\,2 \int_\Delta\,
g(z,z)\;\frac{\partial \big(\mu b^2
e^{\varphi_{\scriptscriptstyle\hspace{-.05cm}B}}\big)}{\partial
\eta_j}\;d^2z\hspace{1.2cm} \,j\,=\,1,\dots,\, N
\end{equation}
where $g(z,t)$ is the Green function on the classical background
described by $\varphi_{\scriptscriptstyle\hspace{-.05cm}B}$ and,
due to the boundary behavior (\ref{varphiB bc infinity}), we have
that
\begin{equation}\label{e^phiB eta derived infinity}
 \frac{\partial\big(\mu b^2
e^{\varphi_{\scriptscriptstyle\hspace{-.05cm}B}}\big)}{\partial
\eta_j} \,=\, O(1) \hspace{.9cm}\textrm{when}\hspace{.4cm}
|z|\rightarrow 1\;.
\end{equation}
This asymptotic behavior can be explicitly checked for the
conformal factor (\ref{phiclassic}) of the $N=1$ case. Formula
(\ref{det integral}) exposes the key role of the Green function at
coincident points. Obviously $g(z,z)$ has to be regularized and we
shall show in what follows that the regularization proposed by
Zamolodchikov and Zamolodchikov (ZZ) in \cite{ZZpseudosphere},
i.e.
\begin{equation}\label{ZZ regularization}
    g(z,z)\,\equiv\,\lim_{t\,\rightarrow\,
    z}\,\left\{\,g(z,t)+\frac{1}{2}\,
\log\left|\,z-t\,\right|^2\,\right\}\;
\end{equation}
gives rise to the correct quantum dimensions. To this end we
examine the transformation properties of the
quantum determinant.\\
From the equation satisfied by the Green function, that is
\begin{equation}\label{g(z,z') equation}
    D\,g(z,z')\,\equiv\,\left(-\,\frac{2}{\pi}\,
\partial_z\partial_{\bar{z}}\,+ \,4\,\mu
    b^2\,e^{\varphi_{\scriptscriptstyle\hspace{-.05cm}B}}\right)\,g(z,z') \,=
\, \delta^2(z-z'\hspace{.05cm})
\end{equation}
we see that, under a $SU(1,1)$ transformation
\begin{equation}\label{SU(1,1) definition}
    z\;\;\longrightarrow\;\; w\,=\,
    \frac{a\,z + b}{\,\rule{0pt}{.4cm}\bar{b}\,z + \bar{a}}
    \hspace{2cm}|a|^2-|b|^2\,=\,1
\end{equation}
$g(z,z'\hspace{.05cm})$ is invariant in value, i.e.
\begin{equation}
    g(z,z')
    \;\;\longrightarrow\;\;\tilde{g}(w,w')\,=\,g(z,z')\;.
\end{equation}
Instead, because of the term $\log\left|z-t\right|$ in (\ref{ZZ
regularization}), the function $g(z,z)$ transforms as follows
under $SU(1,1)$
\begin{equation}\label{g(z,z) SU(1,1) transformation }
    g(z,z)
    \;\;\longrightarrow\;\;\tilde{g}(w,w)\,=\,g(z,z) +
    \,\frac{1}{2}\,\log \left|\,\frac{dw}{dz}\,\right|^2
\end{equation}
where $w(z)$ is given by (\ref{SU(1,1) definition}). Then, the
transformation law for the expression (\ref{det integral}) is
\begin{eqnarray}\label{det transformation}
\frac{\partial}{\partial \eta_j}\,\log
\big(\,\textrm{Det}\,\widetilde{D}\,\big)^{-1/2} & = & -\,2
\int_\Delta\, \tilde{g}(w,w)\;\frac{\partial \big(\mu b^2
e^{\varphi_{\scriptscriptstyle\hspace{-.05cm}B}}\big)}{\partial
\eta_j}\;d^2w\\
\label{det transformation bis} & = & \frac{\partial}{\partial
\eta_j}\,\log \big(\,\textrm{Det}\,D\,\big)^{-1/2} \,-\,
\int_\Delta \log \left|\,\frac{dw}{dz}\,\right|^2\frac{\partial
\big(\mu b^2
e^{\varphi_{\scriptscriptstyle\hspace{-.05cm}B}}\big)}{\partial
\eta_j}\;d^2z \phantom{xxxx}
\end{eqnarray}
where we have used the fact that the $SU(1,1)$ transformation
(\ref{SU(1,1) definition}) does not depend on the charges
$\eta_j$. The Liouville equation (\ref{liouville eq with sources
varphi}) allows to write the second term of (\ref{det
transformation bis}) as follows
\begin{equation}
-\,\frac{1}{2\pi} \,\lim_{r\,\rightarrow\,1}\, \Bigg\{\,
\frac{\partial}{\partial\eta_j}\, \lim_{\varepsilon_n
\,\rightarrow\, 0}\, \int_{\Delta_{\,r,\,\varepsilon}}
\hspace{-.3cm}\log \left|\,\frac{dw}{dz}\,\right|^2
\partial_z\partial_{\bar{z}}\varphi_{\scriptscriptstyle\hspace{-.05cm}B}
\,d^2z\,\Bigg\} 
\end{equation}
where the integration domain is the same occurring in
(\ref{geometric action}). This integral can be computed
integrating by parts. Since $a/b \notin \Delta$, it becomes
\begin{eqnarray}\label{dimension intermediate int}
-\,\frac{\partial}{\partial\eta_j} \,\sum_{k\,=\,1}^N\eta_k\,\log
\left|\,\frac{dw}{dz}\,\right|^2_{z\,=\,z_k}\hspace{-.3cm}  & & \\
& & \hspace{-4cm}+ \lim_{r\,\rightarrow\,1}
\frac{\partial}{\partial\eta_j} \,\left[\,\frac{1}{4\pi i}\;
\oint_{\partial\Delta_{r}}
\varphi_{\scriptscriptstyle\hspace{-.05cm}B}\,\partial_z \log
\frac{dw}{dz}\;dz\,+\,\frac{1}{4\pi i}\;
\oint_{\partial\Delta_{r}}
\partial_{\bar{z}}\varphi_{\scriptscriptstyle\hspace{-.05cm}B}\, \log
\left|\,\frac{dw}{dz}\,\right|^2\hspace{-.1cm}d\bar{z}\,\right]
\rule{0pt}{.84cm}\,\nonumber
\end{eqnarray}
where the first line comes from the circles $\gamma_n$ around the
sources, as $\varphi_{\scriptscriptstyle\hspace{-.05cm}B}$ behaves
like in (\ref{varphiB bc sources}). Then, because of the
asymptotic (\ref{varphiB bc infinity}), the second line of
(\ref{dimension intermediate int}) vanishes in the limit
$r\rightarrow 1$ and we are left only with the first line.\\
By integrating  back, we find
\begin{equation}
\log\big(\,\textrm{Det}\,\widetilde{D}\,\big)^{-1/2} \,=\,
\log\big(\,\textrm{Det}\,D\,\big)^{-1/2}-\sum_{k\,=\,1}^N\eta_k\,\log
\left|\,\frac{dw}{dz}\,\right|^2_{z\,=\,z_k}
\hspace{-.2cm}+f(z_1,\dots,z_N)
\end{equation}
where $f(z_1,\dots,z_N)$ is a function of the positions of the
sources and of the transformation parameters but, since when all
the charges vanish we have
\begin{equation}
\left.\big(\,\textrm{Det}\,\widetilde{D}\,\big)^{-1/2}\,\right|_{\,\eta_i\,
=\,0}\,=\, 
\left.\big(\,\textrm{Det}\,D\,\big)^{-1/2}\,\right|_{\,\eta_i\,=\,0}\,=\,1
\end{equation}
then $f(z_1,\dots,z_N)$ vanishes identically and the
transformation law for the quantum determinant under $SU(1,1)$
reads
\begin{equation}
\big(\,\textrm{Det}\,\widetilde{D}\,\big)^{-1/2} \,=\,
\prod_{n\,=\,1}^N \left|\, \frac{d w}{dz}\, \right|_{z\,=\,z_n}^{-
\,2\eta_n} \hspace{0cm} \big(\,\textrm{Det}\,D\,\big)^{-1/2}\;.
\end{equation}
Comparing this result with (\ref{semiclassic trasf N point}), we
find that the semiclassical dimensions $\eta(1-\eta)/b^2$ are
modified by a quantum correction to
\begin{equation}\label{quantum dimensions}
    \Delta_{\alpha}\,=\,\frac{\eta(1-\eta)}{b^2}\,+ \eta\,=\,
    \alpha(Q-\alpha)
\end{equation}
i.e.
\begin{equation}\label{trasf N point}
\langle \, \widetilde{V}_{\alpha_1}(w_1)\dots
\widetilde{V}_{\alpha_N}(w_N) \,\rangle\,=\, \prod_{n\,=\,1}^N
\left|\frac{d w}{dz} \right|_{z=z_n}^{- \,2\,(\eta_n
(1-\eta_n)/b^2+\eta_n)} \hspace{-.2cm}\langle \,
V_{\alpha_1}(z_1)\dots V_{\alpha_N}(z_N) \,\rangle\;.
\end{equation}
The quantum conformal dimensions (\ref{quantum dimensions}) are
the ones found in \cite{CT} within the hamiltonian approach.\\

\noindent There are no further corrections to the quantum
conformal dimensions (\ref{quantum dimensions}) of the vertex
operator $V_\alpha(z)= e^{2\alpha\phi(z)}$. This statement can be
proved to all order in $b$ by direct inspection of the
perturbative graphs occurring in the expansion in $\alpha$ and
$b$, along the line of
\cite{ZZpseudosphere,MTgeometric,MTtetrahedron}; therefore now the
propagator is given by (\ref{ZZ propagator}).\\
Since we adopt a $SU(1,1)$ non invariant regularization
\cite{ZZpseudosphere,MTgeometric}, the graphs that can modify the
conformal dimensions are only the ones containing tadpoles or
simple loops. Let us consider a vertex which bears $r$ simple
loops, $m-r$ tadpoles and $k$ ordinary propagators, as shown in
the following
figure\\
\[\hspace{-1cm}
\begin{minipage}[c]{4.5cm}
  \includegraphics[width=4.5cm]{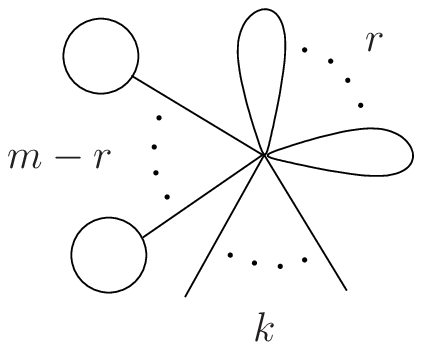}
\end{minipage}
\]
The order of this vertex is $k+2r+(m-r)=k+r+m$. It is generated by
the interaction term
\begin{equation}\label{interactionterm}
\int_{\Delta}
\frac{\big(2b\chi(z)\big)^{k+m+r}}{(k+m+r)!}\;\,d\nu(z)\;
\frac{1}{(m-r)!}\,\prod_{s=1}^{m-r}\left(-
\int_{\Delta}\frac{\big(2b\chi(z_s)\big)^{3}}{3!}\;\,d\nu(z_s)\right)
\end{equation}
where the measure is defined as
\begin{equation}
d\nu(z)\,\equiv\,\mu\,
e^{\left.\varphi_{cl}(z)\right|_{\eta\,=\,0}}\,d^2z\,=\,\frac{d^2z}{\pi
b^2(1-z\bar{z})^2}\;.
\end{equation}
The effective vertex due to (\ref{interactionterm}) is
\begin{eqnarray}\label{general term}
\rule{0pt}{.9cm} \int_{\Delta}
d\nu(z)\,\frac{(2b)^{k+m+r}}{(k+m+r)!}\, {k+m+r \choose k}\,
      \chi^k(z)\,\times \\
 \rule{0pt}{.7cm}     & &\hspace{-2cm}\times\,
{m+r \choose m-r} \,\left(-\frac{2^3
      b^3}{3!}\,3\,P(z)\,\right)^{m-r}
      (2r-1)!!\;\hat g(z,z)^r
     \nonumber
\end{eqnarray}
where $P(z)$ is the tadpole contribution
\begin{equation}
P(z)= \int_{\Delta} {\hat g}(z,z')\;{\hat g}(z',z')\, d\nu(z')\;.
\end{equation}
Adopting the ZZ regularization procedure \cite{ZZpseudosphere},
the propagator at coincident points (i.e. the simple loop) is
given by
\begin{equation}\label{ZZ regularization eta=0}
    \hat g(z,z)\,\equiv\,\lim_{t\,\rightarrow\,
    z}\,\left\{\,\hat g(z,t\hspace{.06cm})+\frac{1}{2}\,
\log\left|\,z-t\,\right|^2\,\right\}
    \,=\,\log(1-z\bar z)-1\;.
\end{equation}
Working out the factorials in (\ref{general term}) and summing
over the number $r$ of simple loops, we have
\begin{eqnarray}
\sum_{r\,=\,0}^m \int_{\Delta} d\nu(z)\;\frac{(2 b
\chi(z))^k}{k!}\; \frac{(2b^2)^m}{m!}\; \big(-4 b^2
P(z)\big)^{m-r} {m \choose m-r}\,
\hat g(z,z)^r\;=\;\\
\rule{0pt}{.8cm} & & \hspace{-7.2cm}=\; \int d\nu(z)\;\frac{(2 b
\chi(z))^k}{k!}\; \frac{(2b^2)^m}{m!}\; \big(-4
 b^2 P(z)+{\hat g}(z,z)\,\big)^m\;.
 \nonumber
\end{eqnarray}
Using the equation for the propagator $\hat g(z,z')$ is easy to
show that \cite{ZZpseudosphere}
\begin{equation}\label{ZZidentity}
-\,4\, b^2\,P(z)\,+\,{\hat g}(z,z) \,=\, \frac{1}{2}\;.
\end{equation}
Notice that if one chooses a $SU(1,1)$ invariant regularization
instead of (\ref{ZZ regularization eta=0}), then the identity
(\ref{ZZidentity}) has a vanishing right hand side.\\
Repeating the argument for all vertices bearing tadpoles and
simple loops, we are left with
a convergent graph which is invariant under $SU(1,1)$.\\
As shown in \cite{MTgeometric}, starting from the regularized
action (\ref{classicalaction}), the exponentiation of the
following graph
\[\hspace{-.6cm}
\vspace{.5cm} \rule{0pt}{1.1cm}
\begin{minipage}[c]{4cm}
  \includegraphics[width=4.5cm]{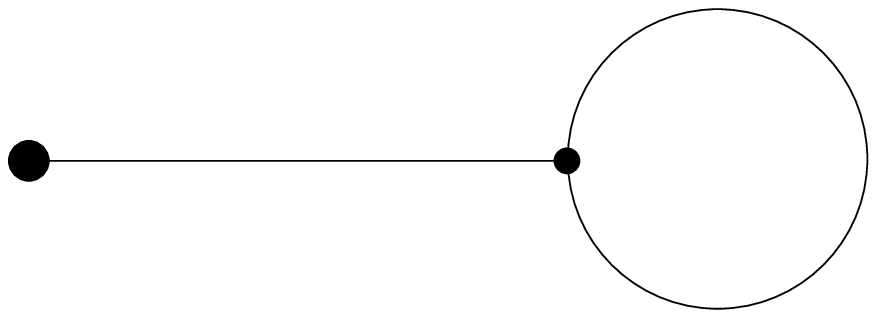}
\end{minipage}
\]
changes the dimensions of the vertex operator $V_\alpha(z)$ from
the semiclassical value $\alpha(1/b-\alpha)$ to the value
$\alpha(1/b+b-\alpha)=\alpha(Q-\alpha)$. We recall
\cite{ZZpseudosphere,MTgeometric} that in the standard approach in
which one simply adds sources to the action \cite{ZZpseudosphere},
the change from the naive dimensions $\alpha/b$ to the
semiclassical dimensions $\alpha(1/b-\alpha)$ is provided by the
exponentiation of the simple loop (\ref{ZZ regularization eta=0}),
which is absent in the approach adopted in \cite{MTgeometric} that
starts from the classical regularized action
(\ref{classicalaction}) and recovers $\alpha(1/b-\alpha)$ at the
semiclassical level.

\section{The Green function on the classical
background}\label{Greenfunctionsect} 

Our next aim will be to compute the exact Green function on the
classical background $\varphi_{cl}$ given in (\ref{phiclassic}).
We shall employ the method developed in \cite{MV}. The procedure
allows also to compute the first term of the expansion in
$\varepsilon$ of the conformal factor in presence of the source in
$z=0$ ($\Delta$ representation) with finite charge $\eta$ and of
another source with infinitesimal charge $\varepsilon$ elsewhere.

\noindent First, we recall that the general solution of the
Liouville equation in presence of $N$ sources is given by
\begin{equation}
\pi \mu b^2 \,e^{\varphi(z)} = \frac{|\,\omega_{12}\,|^2}
    {\rule{0pt}{.6cm}\Big(\,y_1(z)\,\overline{\rule{0pt}{.34cm}y_1(z)}-
    y_2(z)\,\overline{\rule{0pt}{.34cm}y_2(z)}\;\Big)^2}
\end{equation}
where $y_i(z)$ are two independent solutions of the fuchsian
differential equation
\begin{equation}\label{fuchsian eq}
    \frac{d^{\,2} y}{dz^2}\,+ Q(z)\,y\,=\,0
\end{equation}
and $\omega_{12}$ is their wronskian $\omega_{12}=y_1
y_2'-y_1'y_2$. The expression of $Q(z)$ is given by
\begin{equation}
e^{\varphi/2}\,\partial_{z}^2 e^{-\varphi/2} \,=\;
\frac{1}{4}\,(\partial_{z} \varphi)^2-\frac{1}{2}\;\partial_{z}^2
\varphi\;=\, -\, Q(z) \,=\, -\,b^2\, T(z)
\end{equation}
where $T(z)$ is the analytic component of the classical energy
momentum tensor.\\ The analytic function $Q(z)$ contains both
double poles, whose residues are related to the charges $\eta_n$,
and simple poles, whose residues are the Poincar\'e accessory
parameters and have to be determined by imposing the monodromy
condition on the solution.\\
Under a change $z \rightarrow  \xi(z)$ the transformation law of
the solutions of (\ref{fuchsian eq}) is given by
\begin{equation}
y(z)\;\;\;\longrightarrow\;\;\;\tilde{y}(\xi)\,=
\,\big(z'(\xi)\big)^{-1/2}\,y\big(z(\xi)\big)\;. 
\end{equation}
It ensures that the wronskian and the measure
$e^{\varphi(z)}dz\wedge d\bar{z}$ are separately invariant.\\
The energy momentum tensor must satisfy some boundary conditions
guaranteeing that there is neither energy momentum flow
\cite{Cardy,difrancesco} nor singularity at infinity. These
conditions can be formulated in a clearer way in the upper half
plane $\mathbb{H}=\{\,\xi\in\mathbb{C}\,;\,
\textrm{Im}(\xi)>0\,\}$ representation; therefore, for the first
part of the procedure, we shall work in this domain.\\
The Cayley transformation
\begin{equation}
\xi\,=\,-\,i\; \frac{z+1}{z-1}
\hspace{1cm}\longleftrightarrow\hspace{1cm}
z\,=\,\frac{\xi-i}{\xi+i}
\end{equation}
maps the upper half plane $\mathbb{H}$ into the unit disk $\Delta$
and viceversa. Since its
Schwarzian derivative vanishes, we have that
\begin{equation}\label{Q transfom null schwartz}
    Q(z)\,=\,\big(\xi'(z)\big)^{2}\,\widetilde{Q}\big(\xi(z)\big)\,=\,
    -\,\frac{4}{(1-z)^4}\;\widetilde{Q}\big(\xi(z)\big)\;.
\end{equation}
In the upper half plane representation, the condition of no energy
momentum  flow at infinity \cite{Cardy,difrancesco} is that
$\widetilde{T} \,=\,\overline{\rule{0pt}{.44cm}\widetilde{T}}$ on
the real axis, which translates into
\begin{equation}\label{cardy condition}
   \widetilde{Q}(\xi)
   \,=\,\overline{\rule{0pt}{.45cm}\widetilde{Q}}(\xi) \hspace{1.5cm} 
   \textrm{when}\hspace{1.5cm} \xi\in\mathbb{R}
\end{equation}
and therefore, by analyticity, for all $\xi \in \mathbb{H}$.
Instead, the condition of regularity of $\widetilde{Q}(\xi)$ at
infinity is
\begin{equation}\label{regularity condition inf}
   \xi^4\,\widetilde{Q}(\xi) \;\sim\; O(1) \hspace{1.5cm}
   \textrm{when}\hspace{1.5cm}\xi 
   \longrightarrow \infty\;.
\end{equation}

\noindent Let us begin with the unperturbed case of a single
source of finite charge $\eta$. Because of the $SL(2,\mathbb{R})$
invariance of the upper half plane, we can place this source in
$\xi= i$.\\
The function $\widetilde{Q}_0(\xi)$ for the unperturbed case
satisfying (\ref{cardy condition}) can be written as
\begin{equation}\label{Q0 on H}
    \widetilde{Q}_0(\xi)\,=\;\frac{1-\lambda_{i}^2}{4\,(\,\xi-i\,)^2}\,+\,
    \frac{1-\bar{\lambda}_{i}^2}{4\,(\,\xi+i\,)^2}
    \,+\,\frac{b_i}{2\,(\,\xi-i\,)}\,+\,\frac{\bar{b}_{i}}{2\,(\,\xi+i\,)}\;.
\end{equation}
The complex numbers $b_{i}$ and $\bar{b}_{i}=b_{-i}$ are the
unperturbed accessory parameters related to the singularities in
$i$ and in its image $-i$, respectively. The parameter
$\lambda_{i}^2$ is related to the charge $\eta$ as follows
\begin{equation}
   \eta(\eta-1)+ \frac{1-\lambda_{i}^2}{4}\,=\,0
\end{equation}
which tells us that $\lambda_{i}=\bar{\lambda}_{i}$, being $\eta
\in \mathbb{R}$. Moreover, by imposing the regularity condition at
infinity  (\ref{regularity condition inf}) for
$\widetilde{Q}_0(\xi)$, we find
\begin{equation}
    b_{i}\,=\,i\,2\eta(1-\eta)
\end{equation}
and the expression of $\widetilde{Q}_0(\xi)$ becomes
\begin{equation}\label{Q0 on H final}
        \widetilde{Q}_0(\xi)\,=\,\frac{4\eta(\eta-1)}{(\xi^2+1)^2}\;.
\end{equation}
In the $\Delta$ representation, it reads
\begin{equation}
    Q_0(z)\,=\,\frac{\eta(1-\eta)}{z^2}\;.
\end{equation}
Two independent solutions are $y_1(z)=z^{\eta}$ and
$y_2(z)=z^{1-\eta}$ and their wronskian
$\omega_{12}=y_1y_2'-y_1'y_2=1-2\eta$ is constant. Except for a
numerical factor, they correspond respectively to
$\tilde{y}_1(\xi)=(1+i\xi)^{\eta}(1-i\xi)^{1-\eta}$ and
$\tilde{y}_2(\xi)=(1-i\xi)^{\eta}(1+i\xi)^{1-\eta}$ in the
$\mathbb{H}$ representation.\\

\noindent
 Now we perturb the previous geometry by introducing a
 new source at a generic point $\zeta \in\mathbb{H}$ with a
 small charge $\eta_2=\varepsilon$. \\
 We can write down the
 perturbed energy momentum tensor satisfying
(\ref{cardy condition}) as follows
\begin{equation}\label{Q on H}
\widetilde{Q}(\xi)
\,=\,\widetilde{Q}_0(\xi)\,+\,\varepsilon\,\tilde{q}(\xi)
\end{equation}
where $\widetilde{Q}_0(\xi)$ is the unperturbed energy momentum
tensor (\ref{Q0 on H final}) and the perturbation $\tilde{q}(\xi)$
is given by
\begin{equation}\label{q on H}
\tilde{q}(\xi)\,=\, \frac{1}{\rule{0pt}{.4cm}(\,\xi-\zeta\,)^2}
\,+\,\frac{1}{\rule{0pt}{.4cm}(\,\xi-\bar{\zeta}\,)^2}
\,+\,\frac{\beta_{i}}{\rule{0pt}{.4cm}2\,(\,\xi-i\,)}
\,+\,\frac{\bar{\beta}_{i}}{\rule{0pt}{.4cm}2\,(\,\xi+i\,)}
\,+\,\frac{\beta_{\zeta}}{\rule{0pt}{.4cm}2\,(\,\xi-\zeta\,)}
\,+\,\frac{\bar{\beta}_{\zeta}}{\rule{0pt}{.4cm}2\,(\,\xi-\bar{\zeta}\,)}\;.
\phantom{***}
\end{equation}
Notice that now the accessory parameters are given by the sum of
their unperturbed values, already determined, and a perturbation
$O(\varepsilon)$, whose complex coefficients (\,i.e. $\beta_{i}$
and $\beta_{\zeta}$  for the points $i$ and $\zeta$ respectively)
must satisfy the above mentioned conditions.\\
The regularity condition for
 $\xi^3 \tilde{q}(\xi)\rightarrow 0$ when $\xi
\rightarrow \infty$ implies that
\begin{equation}\label{regular infty}
    \left\{\,\begin{array}{l}
\beta_i + \bar{\beta}_i\,+\,\big(\,\beta_\zeta+\bar{\beta}_\zeta\,\big)\,=\,0\\
\rule{0pt}{.6cm}
4-i\,\big(\,\beta_\zeta+\bar{\beta}_\zeta\,\big)\,
+\zeta\,\beta_\zeta+\bar{\zeta}\,\bar{\beta}_\zeta\,=\,0\\
\rule{0pt}{.6cm}4\,\big(\,\zeta+\bar{\zeta}\,\big)\,-\,\big(\,\beta_i+
\bar{\beta}_i\,\big)\, 
+\zeta^2\,\beta_\zeta+\bar{\zeta}^{\,2}\,\bar{\beta}_\zeta\,=\,0\;.
    \end{array}\right.
\end{equation}
We can use a $SL(2,\mathbb{R})$ transformation which leaves $i$
fixed to move the point $\zeta$ on the imaginary axis, $\zeta=
i\tau$, with $\tau\in \mathbb{R}^+_0$ and $\tau\neq 1$.
The system (\ref{regular infty})
simplifies to
\begin{equation}
    \left\{\,\begin{array}{l}
\textrm{Re}(\beta_i)\,=\,\textrm{Re}(\beta_{i\tau})\,=\,0\\
\rule{0pt}{.5cm}
\textrm{Im}(\beta_i)\,=\,2\,-\,\tau\,\textrm{Im}(\beta_{i\tau})\,\equiv\,
\beta
    \end{array}\right.
\end{equation}
and we are left only with the parameter $\beta$ to determine.\\
Through the transformation law (\ref{Q transfom null schwartz}),
we can write the expression $q(z)$ of the perturbation in the
$\Delta$ representation
\begin{equation}\label{q on Delta}
    q(z)\,=\,\frac{1}{\rule{0pt}{.4cm}(\,z-t\,)^2}\,+\,
    \frac{1}{\rule{0pt}{.4cm}(\,z-1/t\,)^2}\,-\,\frac{\beta}{z}\,
    +\left(\,\frac{2\,t+\beta}{\rule{0pt}{.4cm}1-t^2}\,\right)\,
\frac{1}{\rule{0pt}{.4cm}z-t}\,-\, 
    \left(\,t\,\frac{2+t\,\beta}{\rule{0pt}{.4cm}1-t^2}\,\right)\,
\frac{1}{\rule{0pt}{.4cm}z-1/t}
    \phantom{===}
\end{equation}
where
\begin{equation}
    t\,=\,\frac{\tau-1}{\tau+1}\;\;\in\, (-1,1)\setminus\{0\}
\end{equation}
is the image in $\Delta$ of the point $i\tau\in\mathbb{H}$ through
the Cayley transformation.\\

\noindent In the perturbed case, the conformal factor has the
usual structure
\begin{equation}\label{e^phi perturbed}
\pi \mu b^2 \,e^{\varphi_2(z)} = \frac{|\,\Omega_{12}|^2}
    {\Big(\,Y_1(z)\,\overline{\rule{0pt}{.36cm}Y_1(z)}-
    Y_2(z)\,\overline{\rule{0pt}{.36cm}Y_2(z)}\;\Big)^2}
\end{equation}
where $\Omega_{12}=Y_1 Y_2'-Y_1'Y_2$. \\
The solutions $Y_j(z)$ of the perturbed problem can be written as
a sum of the unperturbed solutions
$y_j(z)$ and of a perturbation $O(\varepsilon)$ as follows
\begin{equation}
    Y_i(z)\,=\,y_i(z)\,+\,\varepsilon\,\delta y_i(z) \hspace{2cm} i\,=\,1,\,2
\end{equation}
where $\delta y_j(z)$ satisfy the following inhomogeneous
differential equation
\begin{equation}\label{fuchsian eq deltay}
    \frac{d^{\,2} \delta y_i}{dz^2}\,+ Q_0(z)\,\delta
    y_i\,=\,-\,q(z)\,y_i\;.
\end{equation}
The solutions of this equation are given by the following
integrals \cite{MV}
\begin{eqnarray}
\delta y_i(z) & = & -\,\frac{1}{\omega_{12}}\,\int_{0}^{\,z}
dx\,\Big(y_1(x)\,y_2(z)\,-\,y_1(z)\,y_2(x)\Big)\,q(x)\,y_i(x)\nonumber\\
\rule{0pt}{.8cm}& = &
-\,\frac{1}{\omega_{12}}\,I_{i1}(z)\,y_2(z)\,+\,\frac{1}{\omega_{12}}\,
I_{i2}(z)\,y_1(z)
\end{eqnarray}
where
\begin{equation}\label{Iij(z)}
    I_{ij}(z)\,\equiv\,\int_{0}^{\,z}
    y_{i}(x)\,y_{j}(x)\,q(x)\,dx\;.
\end{equation}
Notice that the integrals $I_{ij}(z)$ are invariant under the
Cayley map
\begin{equation}
    \tilde{I}_{ij}(\xi)\,=\,\int_{i}^{\,\xi}
    \tilde{y}_{i}(y)\,\tilde{y}_{j}(y)\,\tilde{q}(y)\,dy
    \,=\,\int_{0}^{\,z}
    y_{i}(x)\,y_{j}(x)\,q(x)\,dx\,=\,I_{ij}(z)\;.
\end{equation}
Since we have chosen the
position
of the finite source as starting point,
we have that $I_{ij}(0)=\tilde{I}_{ij}(i)=0$.\\
More explicitly, the two independent solutions of the perturbed
problem in terms of the integrals $I_{ij}(z)$ are
\begin{eqnarray}\label{Y solutions}
   Y(z)\,\equiv\, \left(\begin{array}{c}
            Y_1(z)\\
            Y_2(z)
          \end{array}\right)& = &
    \left(\begin{array}{c}
            y_1(z)\\
            y_2(z)
          \end{array}\right)\,+\,\frac{\varepsilon}{\omega_{12}}\,
    \left(\begin{array}{cc}
            I_{12}(z) & -\, I_{11}(z)\\
            I_{22}(z) & -\, I_{12}(z)
          \end{array}\right)
    \left(\begin{array}{c}
            y_1(z)\\
            y_2(z)
          \end{array}\right)\nonumber \\
\rule{0pt}{1cm}& = &
\left(\,\mathbb{I}\,+\,\frac{\varepsilon}{\omega_{12}}\,M_{t}(z)\,\right)
          \left(\begin{array}{c}
            y_1(z)\\
            y_2(z)
          \end{array}\right)\;.
\end{eqnarray}
Notice that, since $\textrm{tr}M_{t}(z)=0$, then
$\Omega_{12}=\omega_{12} + O(\varepsilon^2)$.\\
Moreover, if we consider a finite neighborhood of $z=0$ not
containing $t$ and we let $z$ to
encircle once the origin,
 i.e. $z=\rho\,e^{i\varphi}$
with $0<\rho<t$ with $\varphi$ varying continuously
from $0$ to $2\pi$, then the solutions $Y_1(z)$ and $Y_2(z)$
transform as follows
\begin{equation}
    Y_1(z) \; \longrightarrow \; e^{\,2\pi i \,\eta}\, Y_1(z)
    \hspace{2.6cm}
    Y_2(z) \; \longrightarrow \; e^{\,2\pi i \,(1-\eta)}\,
    Y_2(z)\;.
\end{equation}
This ensures that the conformal factor is monodromic around the
point $z=0$.\\
Now the only freedom left for the vector $Y(z)$ is the
multiplication by the matrix $K \in U(1,1)$
\begin{equation}\label{multiplication freedom}
    Y(z) \,\longrightarrow\, K\,Y(z)
    \hspace{2.5cm}
    K\,=\,\left(\begin{array}{cc}
            k & 0\\
            0 & 1/k
          \end{array}\right)
\end{equation}
where $k=1+\varepsilon\, h(t)$ with $h(t) \hspace{-.1cm}\in
\mathbb{C}$, i.e. the unperturbed value of $k$ must be $1$ in
order to recover the classical solution
(\ref{phiclassic}), which
describes correctly the geometry of the unperturbed case.\\
Now we have to impose the monodromy condition around the point
$z=t$. To do this, we need to compute the change of $I_{ij}(z)$
when $z$ is near the point $t$ and turns once around it. From the
expression of $q(z)$ given in (\ref{q on Delta}), one easily sees
that
\begin{eqnarray}\label{delta Iij}
    \delta I_{ij}(t)&
    = & \oint_{t} \,y_{i}(x)\,y_{j}(x)\,q(x)\,dx \\
  \rule{0pt}{.8cm} & = &  2\pi
    i\,\left(\,\frac{2\,t+\beta}{\rule{0pt}{.4cm}1-t^2}\,\right)\,
    y_i(t)\,y_j(t) + 2\pi i
    \left.\frac{d}{dz}\,\big(\,y_i(z)\,y_j(z)\,\big)\,\right|_{z\,=\,t}\;.
    \nonumber
\end{eqnarray}
Thus, the transformation of the vector $Y(z)$ when one encircles
$z=t$, including also the multiplication (\ref{multiplication
freedom}), is given by the following matrix
\begin{equation}
\mathbb{I} \,+\,\frac{\varepsilon}{\omega_{12}}\,
    \left(\begin{array}{cc}
            \delta I_{12}(t) & -\, \delta I_{11}(t)/k^2\\
            \delta I_{22}(t)/k^2 & -\, \delta I_{12}(t)
          \end{array}\right)\,=\,
\mathbb{I} \,+\,\frac{\varepsilon}{\omega_{12}}\,
    \left(\begin{array}{cc}
            \delta I_{12}(t) & -\, \delta I_{11}(t)\\
            \delta I_{22}(t) & -\, \delta I_{12}(t)
          \end{array}\right)\,+\,O(\,\varepsilon^2)\;.
\end{equation}
The monodromy around $t$ imposes the $U(1,1)$ nature of such a
matrix; therefore, we must require that
\begin{equation}\label{U(1,1) nature}
\overline{\rule{0pt}{.374cm}\delta I_{12}(t)}\,=\,-\,\delta
I_{12}(t)\hspace{2.5cm} \overline{\rule{0pt}{.4cm}\delta
I_{22}(t)}\,=\,-\,\delta I_{11}(t)\;.
\end{equation}
From (\ref{delta Iij}), we have that
\begin{eqnarray}
\delta I_{12}(t)& = & 2\pi
i\left(\,\frac{2\,t+\beta}{\rule{0pt}{.4cm}1-t^2}\;
t+1\,\right)\\
\rule{0pt}{.8cm}\delta I_{11}(t)& = & 2\pi i\;t^{2\eta-1}
\left(\,\frac{2\,t+\beta}{\rule{0pt}{.4cm}1-t^2}\;t+2\,\eta\,\right)\\
\rule{0pt}{.8cm}\delta I_{22}(t)& = & 2\pi
i\;t^{1-2\eta}\left(\,\frac{2\,t+\beta}{\rule{0pt}{.4cm}1-t^2}\;t+2-2\,\eta\,
\right)\;.
\end{eqnarray}
Since $\beta \in \mathbb{R}$, the first condition of (\ref{U(1,1)
nature}) is already realized. Instead, the second condition
provides the explicit expression of $\beta$
\begin{equation}\label{beta}
    \beta \,=\,
-\,2\;\frac{\eta+(1-\eta)\,t^2-t^{2\,(1-2\eta)}\left(1-\eta+\eta\,t^2\right)}{
\rule{0pt}{.4cm}t\,\big(1-t^{2\,(1-2\eta)}\big)}\;.
\end{equation}
Thus, the reflection condition (\ref{cardy condition}) together
with the regularity requirement at infinity (\ref{regularity
condition inf}) and the monodromy of the perturbed solution
(\ref{e^phi perturbed}) around $0$ and $t$
fix completely the perturbed accessory parameters.\\

\noindent Now, if we write $\varphi_{2}(z)$ in (\ref{e^phi
perturbed}) as follows
\begin{equation}\label{phi2classic}
    \varphi_{2}(z)\,=\,\varphi_{cl}(z)+ \epsilon\,
    \psi(z,t)+ O(\epsilon^2)
\end{equation}
where $\varphi_{cl}(z)$ is given by (\ref{phiclassic}), by using
the expression (\ref{Y solutions}) for the perturbed solutions, we
find that \cite{MV}
\begin{eqnarray}\label{psi(z,t)}
 \psi(z,t\hspace{.05cm}) \hspace{-.1cm}& = &\hspace{-.1cm}
 -\,\frac{2}{w_{12}}\, \left\{\; 
\frac{y_1\bar{y}_1+y_2\bar{y}_2}{y_1 \bar{y}_1-y_2\bar{y}_2}\,
\Big(\,I_{12}+\bar{I}_{12}+2\,\textrm{Re}\,h(t)\,\Big)\right.
\\
\rule{0pt}{1cm} & & \hspace{4.3cm}\left. -\;\frac{ \bar{y}_1
y_2\,I_{11}+y_1\bar{y}_2\,I_{22}}{y_1 \bar{y}_1-y_2\bar{y}_2}\,
 -\, \frac{ y_1\bar{y}_2
\,\bar{I}_{11}+\bar{y}_1 y_2\,\bar{I}_{22}}{y_1
\bar{y}_1-y_2\bar{y}_2} \;\right\}\;.\nonumber
\end{eqnarray}
The parameter $\textrm{Re}\,h(t)$ appearing in this expression
cannot be determined through monodromy arguments because the term
it multiplies
\begin{equation}
    f(z)\,\equiv\,\frac{y_1\bar{y}_1+y_2\bar{y}_2}{y_1 \bar{y}_1-y_2\bar{y}_2}
\end{equation}
is a monodromic solution of the homogeneous
differential equation
\begin{equation}\label{liouville eq 2 point}
-\,\partial_z \partial_{\bar{z}}\,f(z)+ 2\pi \,\mu b^2\,
e^{\varphi_{cl}}f(z)\,=\,0\;.
\end{equation}
Instead, $\textrm{Re}\,h(t)$ is fixed through the analysis of the
behavior of $\varphi_{2}$ when $|z| \rightarrow 1$. Indeed, $f(z)$
violates the asymptotic (\ref{varphiB bc infinity}) because it
diverges as $O\big(1/(1-z\bar{z})\big)$ when $|z| \rightarrow 1$;
therefore $\textrm{Re}\,h(t)$ is uniquely determined by imposing
the boundary conditions (\ref{varphiB bc infinity}) for the
classical field $\varphi_{2}$. Since the leading logarithmic
divergence in (\ref{varphiB bc infinity}) is already recovered by
$\varphi_{cl}$, then $\psi(z,t)$ must not diverge when $|z| \rightarrow 1$.\\
\noindent Before computing $\psi(z,t)$ explicitly and examining
its boundary behavior, we notice that $\psi(z,t)$ provides also
the Green function on the classical background with one finite
source, i.e. $\varphi_{cl}$. Indeed, since $\varphi_{2}$ describes
the classical background of the pseudosphere with one finite
source of charge $\eta_1=\eta$ in $z_1=0$ and another source of
infinitesimal charge $\eta_2=\epsilon$ placed in $z_2=t$, it
satisfies the following Liouville equation
\begin{equation}\label{liouville eq 2 point}
-\,\partial_z \partial_{\bar{z}}\varphi_{2} + 2\pi \mu b^2\,
e^{\varphi_{2}}\,=\,2\pi\,\eta \,\delta^2(z) + 2\pi\,\epsilon
\,\delta^2(z-t\,)\;.
\end{equation}
Taking the derivative of this equation w.r.t. $\epsilon$ and
setting $\epsilon=0$, we find that $\psi(z,t)$ solves the
following equation
\begin{equation}\label{eq for psi}
    -\,\partial_z \partial_{\bar{z}}\,\psi +
2\pi \mu b^2 \,e^{\varphi_{cl}}\psi\,= \,2\pi\,\delta^2(z-t\,)
\end{equation}
and therefore the Green function $g(z,t)$ arising from the
quadratic part of the quantum action (\ref{quantum action final})
is given by
\begin{equation}
g(z,t)\,=\,\langle\,\chi(z)\chi(t)\,\rangle\,=\,\frac{1}{4}\;\psi(z,t)\;.
\end{equation}
To understand better the final expression for $g(z,t)$ it more
useful to write it in the form given below
\begin{eqnarray}\label{g(z,t)}
 g(z,t)\hspace{-.2cm} & = &\hspace{-.2cm}  -\,\frac{1}{2w_{12}}
 \,\left\{\;
\frac{y_1\bar{y}_1+y_2\bar{y}_2}{y_1 \bar{y}_1-y_2\bar{y}_2}\;
\Big(\,I_{12}+\bar{I}_{12}+2\,\textrm{Re}\,h(t)\,\Big)\right.
\\
\rule{0pt}{1cm} & & \hspace{1.3cm}\left.
 - \,\frac{ y_1 \bar{y}_1}{y_1 \bar{y}_1-y_2\bar{y}_2}\,
\left(\,\frac{y_2}{y_1}\,I_{11}+\frac{\bar{y}_2}{\bar{y}_1}\,
\bar{I}_{11}\right)
 - \, \frac{ y_2 \bar{y}_2}{y_1 \bar{y}_1-y_2\bar{y}_2}\,
\left(\,\frac{y_1}{y_2}\,I_{22}+\frac{\bar{y}_1}{\bar{y}_2}\,
\bar{I}_{22}\right)\;\right\}\,.
\nonumber
\end{eqnarray}
As noticed before and computed in appendix \ref{details green
function}, $\textrm{Re}\,h(t)$ is fixed by the asymptotic behavior
of $g(z,t)$ when $|z| \rightarrow 1$ and the result is
\begin{equation}\label{Reh(t)}
\textrm{Re}\,h(t) \,=\,
\frac{1}{2}\,\left(\,\frac{1+t^{2\,(1-2\eta)}}{\rule{0pt}{.4cm}
1-t^{2\,(1-2\eta)}}\,\log t^{2\,(1-2\eta)}+2\right)\;.
\end{equation}
Moreover, by exploiting the invariance under rotation, one can
easily generalize all these expressions to a complex $t\in
\Delta$.\\
The Green function in the explicit symmetric form is given by
\begin{eqnarray}\label{g(z,t) symmetric}
 \rule{0pt}{1cm}g(z,t) & = &
 -\,\frac{1}{2}\;
\frac{1+(z\bar{z})^{1-2\eta}}{1-(z\bar{z})^{1-2\eta}}\;
\frac{1+(t\bar{t}\hspace{.04cm})^{1-2\eta}}{\rule{0pt}
{.4cm}1-(t\bar{t}\hspace{.04cm})^{1-2\eta}}\;
\log\omega(z,t\hspace{.04cm})\, -\,\frac{1}{1-2\eta}\,
\\
\rule{0pt}{1cm} & & \hspace{-1.6cm}-\,
\frac{1}{\rule{0pt}{.4cm}1-(z\bar{z})^{1-2\eta}}\;
\frac{1}{\rule{0pt}{.4cm}\,1-(t\bar{t}\hspace{.04cm})^{1-2\eta}}\;\left\{\,
(z\bar{t}\hspace{.05cm})^{1-2\eta}\,\Big(\,B_{\,z/\,t}\big(2\eta,
\,0\big)-B_{\,z\bar{t}\,}\big(2\eta,\,0\big)\,\Big)
\rule{0pt}{.6cm}\right.
\nonumber \\
& &\rule{0pt}{.6cm} \hspace{4cm}\left.
+\,(\bar{z}t\hspace{.04cm})^{1-2\eta}\,\Big(\,B_{\,t/z}\big(2\eta,
\,0\big)-B_{\,1/(z\bar{t}\hspace{.01cm})\,}\big(2\eta,\,0\big)\,\Big)
+\textrm{c.c.}\rule{0pt}{.6cm}\;\right\} \nonumber
\end{eqnarray}
where $\omega(z,t)$ is the $SU(1,1)$ invariant
\begin{equation}\label{SU(1,1) invariant ratio t complex}
    \omega(z,t)\,=\,\left|\,\frac{z-t}{\rule{0pt}{.4cm}1-
    z\,\bar{t}}\,\right|^2
\end{equation}
which is related to the geodesic distance on the pseudosphere
without sources.\\
Notice that only the special case $B_{x}(a,0)$ of the incomplete
beta function $B_{x}(a,b)$ occurs; it is related to the
hypergeometric function $F(a,1;a+1;\,x)$ as follows
\begin{equation}
    B_{x}(a,0)\,=\,\frac{x^a}{a}\,F(a,1;a+1;\,x)\,=\,\int_0^{\,x}
    \frac{y^{a-1}}{1-y}\;dy\,=\,\sum_{n\,\geqslant\; 0}\frac{x^{a+\,n}}{a+n}\;.
\end{equation}
In appendix \ref{details green function} it is shown that $g(z,t)$
is regular at the origin and, through a partial wave expansion, it
is also shown that
\begin{equation}\label{g(z,t) at |z|=1}
g(z,t)\,=\,O\big((1-z\bar{z})^2\big)
\hspace{.9cm}\textrm{when}\hspace{.8cm} |z|\rightarrow 1\;.
\end{equation}
Because of this asymptotic at infinity,  the quantum action
(\ref{quantum action final}) becomes
\begin{equation}\label{quantum action no limit}
     S_{q}[\, \chi\,]
     \,=\,
 \int_{\Delta} \left( \,\frac{1}{\pi} \,\partial_z
\chi \,\partial_{\bar{z}}\chi+2\mu b^2\,
e^{\,\varphi_{cl}}\,\chi^2 \,\right)d^2z\,+\, \sum_{k\,
\geqslant\; 3} \frac{(2b)^k}{k!}\int_\Delta
\mu\,e^{\varphi_{cl}}\,\chi^k\,d^2z
\end{equation}
where $\varphi_{cl}$ is given by (\ref{phiclassic}).\\
A related function that will play a crucial role in what follows
is the Green function (\ref{g(z,t) symmetric}) at coincident
points regularized according to the ZZ procedure, i.e. (\ref{ZZ
regularization}).\\
By computing (\ref{ZZ regularization}) or using the series
representation (\ref{F21 psi 
serie}) of the  
hypergeometric function $F(a,1;1+a;\,x)$, we find that $g(z,z)$ is
given by
\begin{eqnarray}\label{g(z,z)}
\hspace{-0cm} g(z,z) & = &
\Bigg(\,\frac{1+(z\bar{z})^{1-2\eta}}{\rule{0pt}
{.4cm}1-(z\bar{z})^{1-2\eta}}\,\Bigg)^2
\log(1-z\bar{z})\,
-\,\frac{1}{1-2\eta}\;\frac{1+(z\bar{z})^{1-2\eta}}
{\rule{0pt}{.4cm}1-(z\bar{z})^{1-2\eta}}
\\
\rule{0pt}{1cm}
 & &   +\,\frac{2\,(z\bar{z})^{1-2\eta}}{
\big(\rule{0pt}{.45cm}1-(z\bar{z})^{1-2\eta}\big)^2}\, \left(
B_{z\bar{z}}\big(2\eta\,,0\big)+B_{z\bar{z}}\big(2-2\eta\,,0\big)
\rule{0pt}{.5cm}\right.\nonumber\\
& &\hspace{5.45cm} \left.\rule{0pt}{.5cm}+\,
2\gamma_{\scriptscriptstyle\hspace{-.02cm}E}+\psi(2\eta)+\psi(2-2\eta)-\log
z\bar{z} \,\right)\nonumber
\end{eqnarray}
where $\gamma_{\scriptscriptstyle\hspace{-.02cm}E}$ is the Euler
constant and
$\psi(x)=\Gamma'(x)/\Gamma(x)$.\\
The asymptotic behavior at infinity (i.e. when $|z|\rightarrow 1$)
of $g(z,z)$ is the following
\begin{equation}\label{g(z,z) at infinity}
g(z,z)\,=\,\log(1-z\bar{z})-1
-\frac{\eta(1-\eta)}{6}\,(1-z\bar{z})^2+O\big((1-z\bar{z})^3\big)\;.
\end{equation}
Notice that the dependence on the charge $\eta$ occurs only at
$O\big((1-z\bar{z})^2\big)$.

\section{The one point function: the quantum determinant}
\label{quantum determinant sec}

\noindent In this section we compute the quantum determinant for
$N=1$ explicitly and we compare this result with the corresponding
order in the expansion of the one point function
obtained in the bootstrap approach.\\
To compute the quantum determinant for the one point function, we
apply the formula (\ref{det integral}) with
$\varphi_{\scriptscriptstyle\hspace{-.05cm}B}=\varphi_{cl}$ and
with $g(z,z)$ given by (\ref{g(z,z)}), i.e.
for a single source $\eta_1=\eta$ at $z_1=0$.\\
To perform this integral, it is more convenient to adopt the
variable $u\equiv (z\bar{z})^{1-2\eta}$ in the radial integration.
Then, after an integration by parts, we obtain
\begin{equation}\label{log det integral}
\frac{\partial}{\partial \eta}\,\log
\big(\,\textrm{Det}\,D(\eta,0)\,\big)^{-1/2}
=\;2\,\gamma_{\,\scriptscriptstyle\hspace{-.05cm}E}\, +\,2\,
\psi(1-2\eta)\,+\,\frac{3}{1-2\eta}\;.
\end{equation}
Integrating back in $\eta$ with the initial condition
\begin{equation}
\log
\left.\big(\,\textrm{Det}\,D(\eta,0)\,\big)^{-1/2}\,\right|_{\,\eta\,=\,0}\,
=\,0 
\end{equation}
we find the explicit expression of the logarithm of the quantum
determinant
\begin{equation}\label{quantum det eta}
\log \big(\,\textrm{Det}\,D(\eta,0)\,\big)^{-1/2}\,=\,
2\eta\,\gamma_{\,\scriptscriptstyle\hspace{-.05cm}E}-\log
\Gamma(1-2\eta) -\frac{3}{2}\,\log(1-2\eta)\;.
\end{equation}
Putting this result together with the classical contribution
(\ref{one point classical term}), we have the first two terms of
the perturbative expansion in the coupling constant $b$ of the one
point function
\begin{eqnarray}\label{one point serie b}
\rule{0pt}{.6cm} \left\langle \, V_{\eta/b}(0) \,\right\rangle
\;=\;\left\langle \, e^{2(\eta/b)\phi(0)} \,\right\rangle\;=
 & &\hspace{-.45cm}
\exp\left\{-\,\frac{1}{b^2}\;\Big[\,\eta\,\log(\pi \mu
b^2)+2\eta+(1-2\eta)\,\log(1-2\eta)\,\Big]\,\right\}
        \nonumber\\
 \rule{0pt}{.9cm} & & \hspace{-.45cm}
 \times\;\frac{e^{2\eta\gamma_{\,\scriptscriptstyle\hspace{-.05cm}E}}}
{\Gamma(1-2\eta)\,(1-2\eta)^{3/2}}
 \;\big(\,1+O(b^2)\,\big)\;.
\end{eqnarray}
The expansion (\ref{one point serie b}) agrees with the expansion
in the coupling constant of the logarithm of the formula
$U_{1,1}(\eta/b)$ given in (\ref{bootstrap ZZ onepoint}), found by
ZZ \cite{ZZpseudosphere} through the bootstrap method.\\

\noindent We remark that the result (\ref{one point serie b})
corresponds to the summation of two infinite classes of
perturbative graphs computed on the regular background, i.e. with
the classical field given by
$\left.\varphi_{cl}\right|_{\,\eta\,=\,0}$ and the propagator
$\hat g(z,z')$ written in (\ref{ZZ propagator}).
These graphs can be recovered by expanding the logarithm of (\ref{one
  point serie b}) in $\eta=\alpha\, b$.\\ 
The classical part gives rise to the following series of graphs
\begin{eqnarray}
 -\,\frac{1}{b^2}\;\Big[\,\eta\,\log(\pi \mu b^2)+2\eta+(1-2\eta)\,
\log(1-2\eta)\,\Big] & = &
\frac{\eta}{b^2}\,\left.\varphi_{cl}(0)\right|_{\,\eta\,=\,0}-\,2\,
\frac{\eta^2}{b^2}\\
\rule{0pt}{1cm}& & \hspace{-7.5cm}\vspace{.4cm}
\,+\,\frac{\eta^3}{b^2}\, \left\{\;
\begin{minipage}[c]{1cm}
  \includegraphics[width=1cm]{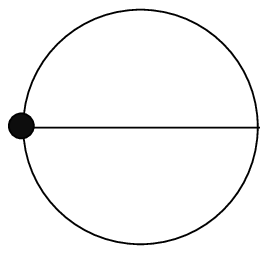}
\end{minipage}\hspace{0cm}\;\right\}\,+\,
\frac{\eta^4}{b^2}\, \left\{\;
\begin{minipage}[c]{2.6cm}
  \includegraphics[width=2.6cm]{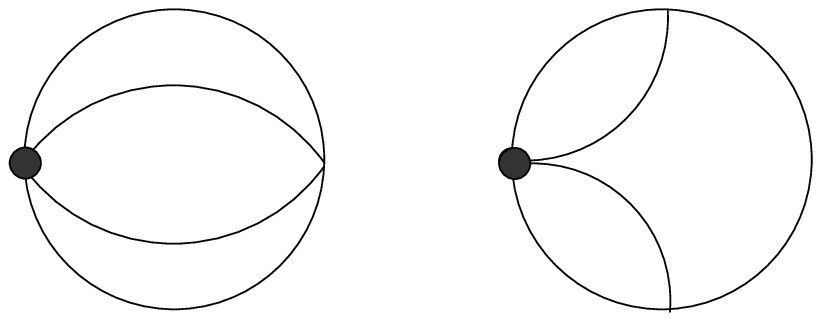}
\end{minipage}\hspace{0cm}\;\right\}
\,+\,\frac{\eta^5}{b^2}\,
\left\{\;
\begin{minipage}[c]{4cm}
  \includegraphics[width=4cm]{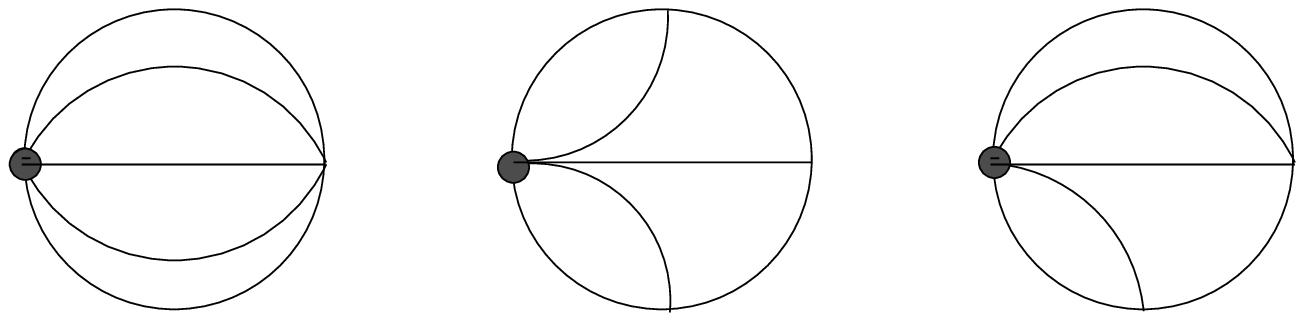}
\end{minipage}\hspace{0cm}\;\right\}
\nonumber\\
\rule{0pt}{1cm} & &\hspace{-7.45cm} +\;\frac{\eta^6}{b^2}\,
\left\{\;
\begin{minipage}[c]{7cm}
  \includegraphics[width=7cm]{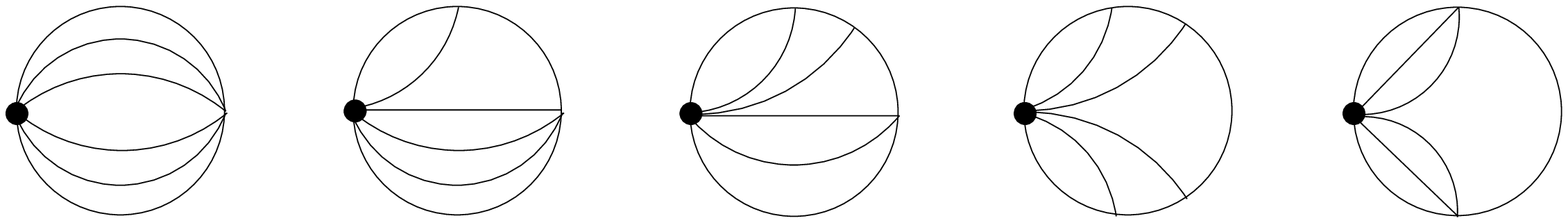}
\end{minipage}\hspace{0cm}\;\right\}\,
+\,\dots\nonumber\\
\rule{0pt}{1cm} & &\hspace{-8.15cm}
=\;\;-\,\frac{\eta}{b^2}\,\log(\pi\mu b^2)\,
-\,2\,\frac{\eta^2}{b^2}\,-\,\frac{4}{3}\;\frac{\eta^3}{b^2}\,
-\,\frac{4}{3}\;\frac{\eta^4}{b^2}\,
-\,\frac{8}{5}\;\frac{\eta^5}{b^2}\,
-\,\frac{32}{15}\;\frac{\eta^6}{b^2}\,+\,\dots\nonumber
\end{eqnarray}
\rule{0pt}{.6cm} while the quantum determinant contribution
contains the following perturbative orders
\begin{eqnarray}
\vspace{.4cm} 2\,\gamma_{\,\scriptscriptstyle\hspace{-.05cm}E}
\,\eta-\log \Gamma(1-2\eta) -\frac{3}{2}\,\log(1-2\eta) & = &
\eta\, \left\{\;
\begin{minipage}[c]{1.7cm}
  \includegraphics[width=1.7cm]{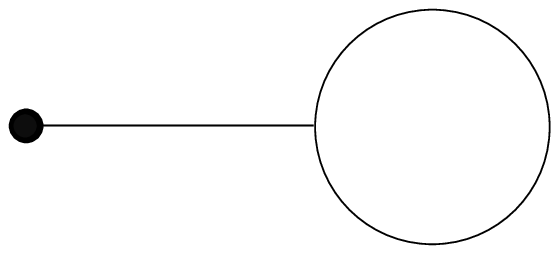}
\end{minipage}\hspace{0cm}\;\right\}  \nonumber \\
\rule{0pt}{1.7cm} & &  \hspace{-6.7cm} + \;\eta^2 \,\left\{\;
\begin{minipage}[c]{4.8cm}
  \includegraphics[width=4.8cm]{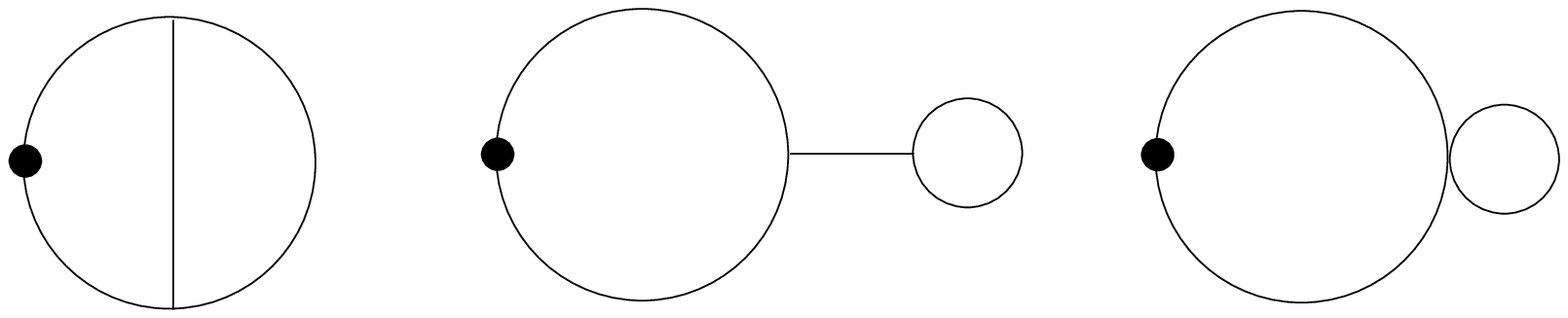}
\end{minipage}\hspace{0cm}\;\right\}  \;+ \;\eta^3\, \left\{\;
\begin{minipage}[c]{5.9cm}
  \includegraphics[width=5.9cm]{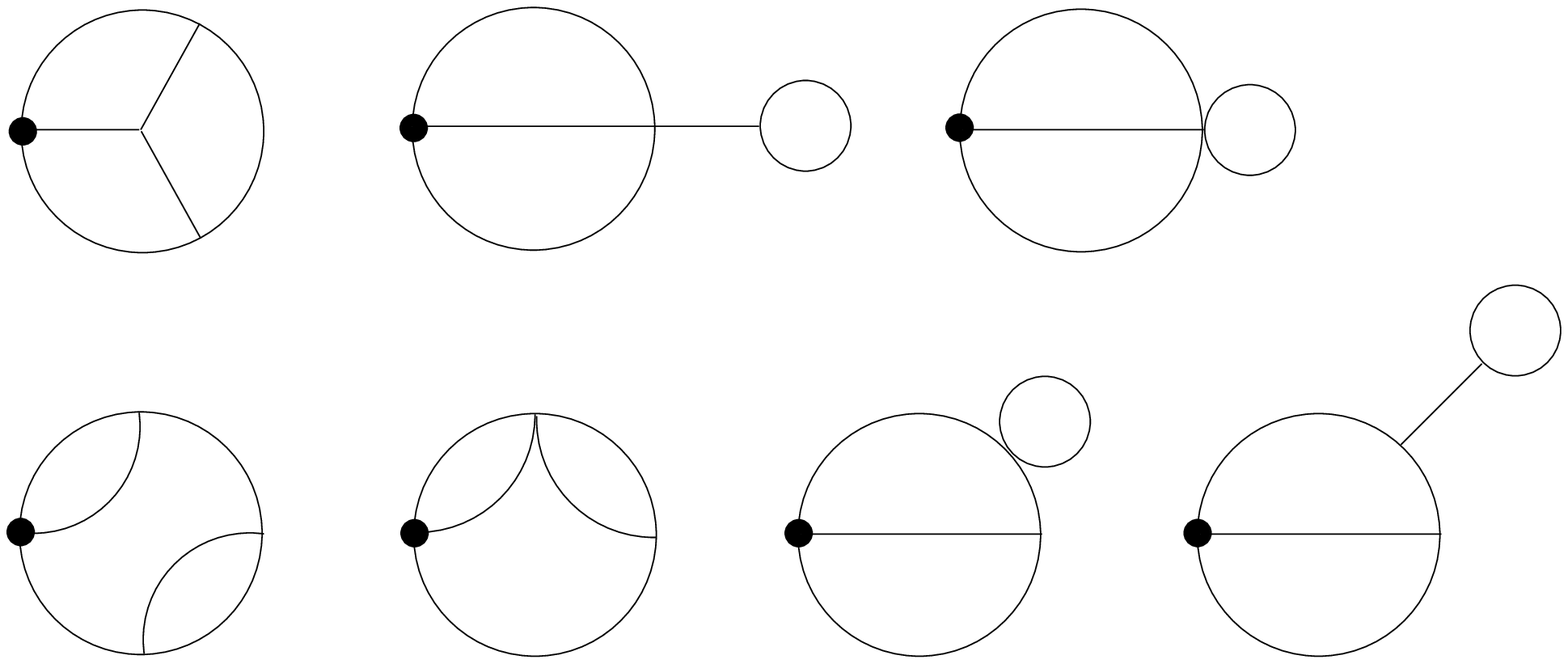}
\end{minipage}\hspace{0cm}\;\right\}\nonumber \\
\rule{0pt}{3.7cm} & & \hspace{-6.7cm} +\;\eta^4 \,\left\{\;\;
\begin{minipage}[c]{9.8cm}
  \includegraphics[width=9.8cm]{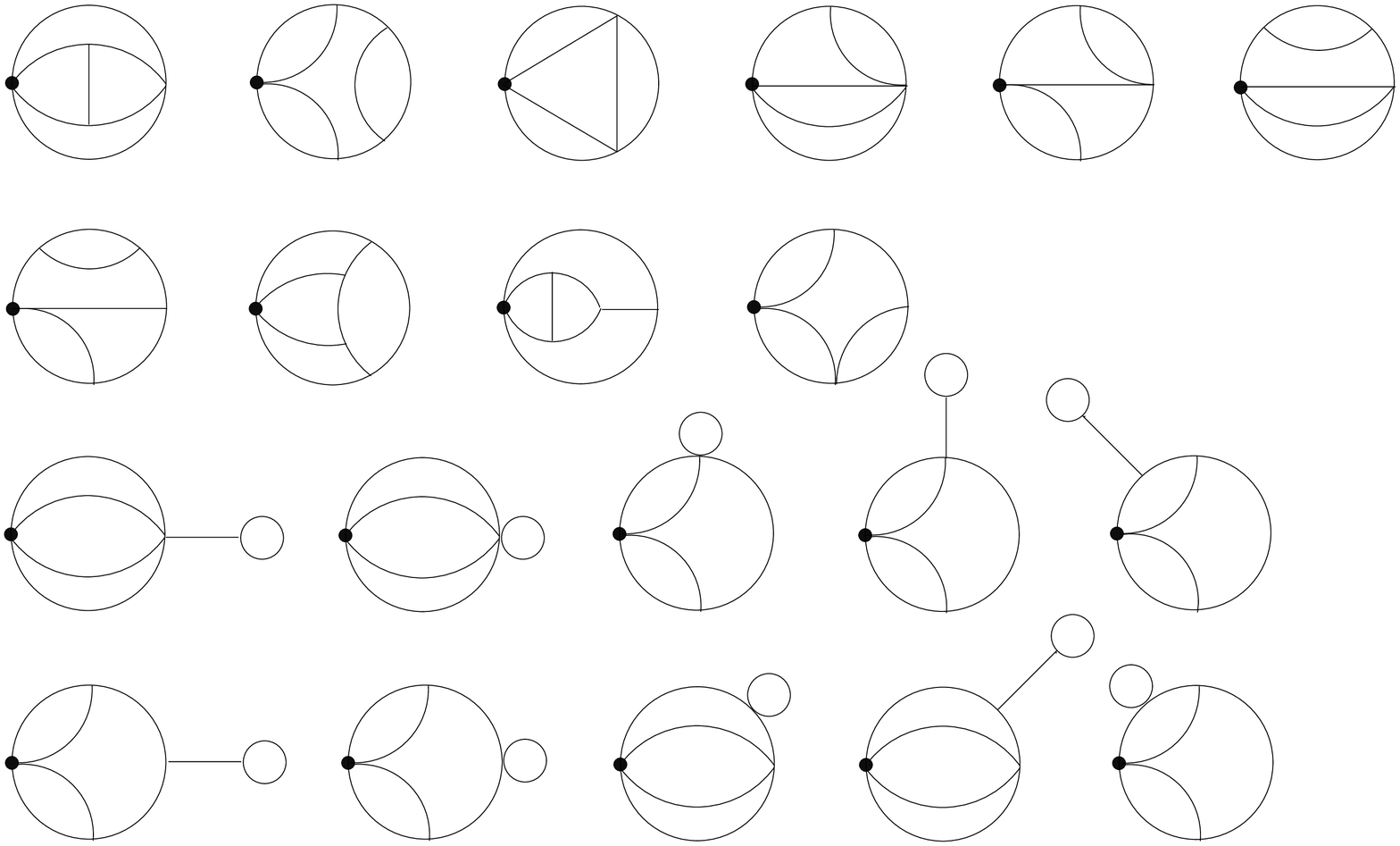}
\end{minipage}\hspace{0cm}\;\;\right\}\;+\;\dots  \\
\rule{0pt}{1.2cm} & & \hspace{-7.5cm} = \;\; 3\,\eta
+\left(3-\frac{\pi^2}{3}\right)\eta^2
+\frac{4}{3}\big(3-2\,\zeta(3)\big)\,\eta^3+
2\left(3-\frac{\pi^4}{45}\right)\eta^4+\,\dots \nonumber
\end{eqnarray}
All the propagators in the figures are given by $\hat{g}(z,z')$.
Those starting from the source were denoted in \cite{MTgeometric}
by dotted lines because they still represent a classical field
even though computationally they are given by the same
expression.\\
The first orders of the classical part have been determined in
\cite{ZZpseudosphere} while the orders $O(\eta^4/b^2)$ and
$O(\eta^5/b^2)$ have been computed in \cite{MTgeometric}. As for
the quantum determinant, the $O(\eta^2)$ contribution agrees with
the result obtained by ZZ \cite{ZZpseudosphere}, while the
$O(\eta^3)$ term agrees with the explicit, but far more difficult
computation performed in \cite{MTtetrahedron}. Instead, the
$O(\eta^4)$ term and the further orders in the quantum determinant
are new results, obtained as byproducts of the knowledge of the
quantum determinant for every value of $\eta <1/2$.\\

\noindent With some effort, one could compute the $O(b^2)$
contribution in (\ref{one point serie b}) within our framework. It
is given by the following three graphs
\[\hspace{-6.3cm}
\rule{0pt}{1.8cm}
\begin{minipage}[c]{4cm}
  \includegraphics[width=10.5cm]{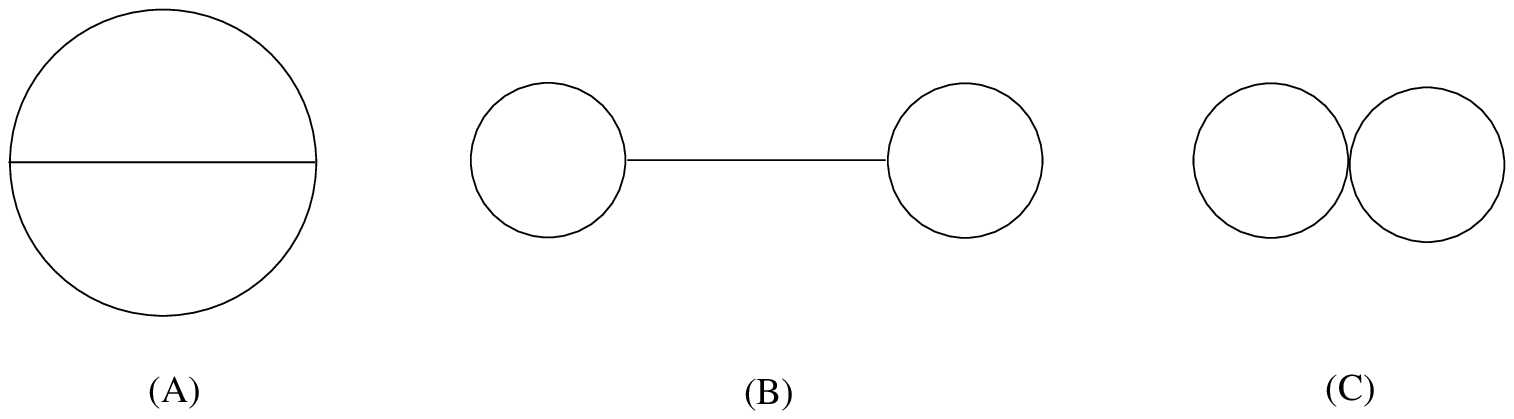}
\end{minipage}
\]
where the propagator is given by (\ref{g(z,t) symmetric}).

\section{The two point function}
\label{two point sec}

\noindent In this section we apply the technique developed in the
previous sections to compute the following two point function on
the pseudosphere
\begin{equation}\label{2 point vertex eta-epsilon}
\left\langle \, V_{\eta/b}(z_1)\,
V_{\varepsilon/b}(z_2)\,\right\rangle
\end{equation}
up to $O(\varepsilon)$ and $O(b^0)$ included, but
to all orders in $\eta$ and in the invariant distance.\\
According to \cite{ZZpseudosphere}, this result is related to the
conformal block with null intermediate dimension through to the
``boundary'' representation of the ``normalized'' two point
function
\begin{equation}\label{conformal block}
    g_{\,\eta/b,\,\varepsilon/b}(\omega)\,\equiv\,\frac{\left\langle 
\, V_{\eta/b}(z_1)\,
V_{\varepsilon/b}(z_2)\,\right\rangle}{\rule{0pt}{.41cm}\left\langle
\, V_{\eta/b}(z_1)\,\right\rangle \left\langle \,
 V_{\varepsilon/b}(z_2)\,\right\rangle}\,=\,(1-\omega)^{2\Delta_{\eta/b}}
 \, \mathcal{F}\left(
 \begin{array}{cc}
\eta/b & \varepsilon/b\\
\eta/b & \varepsilon/b
 \end{array};\,iQ/2,\,1-\omega \right)
\end{equation}
where $\omega(z_1,z_2)$ is the $SU(1,1)$ invariant given in
(\ref{SU(1,1) invariant ratio t complex}).\\
The procedure will be to compute the classical action and the
quantum determinant on the background (\ref{phi2classic})
describing the pseudosphere with two curvature singularities: a
finite one $\eta_1=\eta$ in $z_1=0$ and an infinitesimal one
$\eta_2=\varepsilon$ in $z_2=t$. Since this classical background
is known up to $O(\varepsilon)$, our results will be exact in
$\eta$ and perturbative in $\varepsilon$ up to $O(\varepsilon)$
included. This perturbative background has been already computed
in section \ref{Greenfunctionsect} and it is given by
\begin{equation}\label{varphi2 with g(z,t)}
    \varphi_{2}(z)\,=\,\varphi_{cl}(z)+ 4\,\epsilon
    \,g(z,t)+ O(\epsilon^2)
\end{equation}
where $\varphi_{cl}(z)$ is the background field (\ref{phiclassic})
describing the pseudosphere with a single finite source
$\eta_1=\eta$ placed in $z_1=0$ and $g(z,t)$ is the propagator
(\ref{g(z,t) symmetric}).\\
The two point function (\ref{2 point vertex eta-epsilon}) up to
$O(b^2)$ is
\begin{eqnarray}\label{2 point b serie }
\left\langle \, V_{\eta/b}(0)\,V_{\varepsilon/b}(t)\,\right\rangle
&  = &  e^{-S_{cl}(\eta,\,0;\,\varepsilon,\,t)+S_{cl}(0)} \,\times \\
\rule{0pt}{.9cm}& &\hspace{-1.8cm} \times\;\frac{1}{Z_0}
\int\hspace{-.06cm}\mathcal{D}\,[\, \chi \,]\;\,
\exp\left\{-\,\frac{1}{2}\,\int_\Delta \chi
\left(-\,\frac{2}{\pi}\,\partial_z\partial_{\bar{z}}\,+ \,4\mu
    b^2\,e^{\varphi_{2}}\right)\chi\,d^{2}z\,\right\}
    \big(\,1+O(b^2)\,\big)\nonumber\\
    &   &
\rule{0pt}{.8cm} \hspace{-4cm}=\;\;
e^{-S_{cl}(\eta,\,0;\,\varepsilon,\,t)+S_{cl}(0)}\, \times
\nonumber\\
    &  &
\rule{0pt}{.9cm} \hspace{-3.15cm}
\times\,\big(\,\textrm{Det}\,D(\eta,\,0)\,\big)^{-1/2}\left(\,1\,
-\,8\,\mu b^2\varepsilon \int_{\Delta}g(z,t)\,e^{\varphi_{cl}(z)}
g(z,z)\,d^2z\,+O(\varepsilon^2)\,\right)\big(\,1+O(b^2)\,\big)
\nonumber
\end{eqnarray}
where $S_{cl}(\eta,0;\varepsilon,t)$ is the classical action
(\ref{action classical phiB}) evaluated on the field
$\varphi_{2}(z)$,  while $S_{cl}(0)$ and $Z_0$ are respectively
the classical contribution and the quadratic part of the partition
function $Z$ occurring in (\ref{N point geometric})
and (\ref{partition function}).\\
The denominator occurring in (\ref{conformal block}) with $z_1=0$
and $z_2=t$ up to $O(b^2)$ reads
\begin{equation}
e^{-S_{cl}(\eta,\,0)+S_{cl}(0)}
\big(\textrm{Det}\,D(\eta,\,0)\big)^{-1/2}
\;e^{-S_{cl}(\varepsilon,\,t)+S_{cl}(0)}
\big(\textrm{Det}\,D(\varepsilon,\,t)\big)^{-1/2}
\,\big(\,1+O(b^2)\,\big)
\end{equation}
where $\big(\textrm{Det}\,D(\eta,\,0)\big)^{-1/2}$ has been
already computed and it is given by (\ref{quantum det eta}).\\
Evaluating the classical action (\ref{action classical phiB}) on
the field (\ref{varphi2 with g(z,t)}), we get the following
perturbative expression in $\varepsilon$
\begin{equation}
S_{cl}(\eta,0;\,\varepsilon,t)\,=\,
S_{cl}(\eta,0)-\frac{\varepsilon}{b^2}\,\varphi_{cl}(t)+O(\varepsilon^2)\;.
\end{equation}
This formula for $\eta=0$ provides
 \begin{equation}
S_{cl}(\varepsilon,t)\,=\,S_{cl}(0)
-\,\frac{\varepsilon}{b^2}\left.\varphi_{cl}(t)\right|_{\,\eta\,=
\;0}+O(\varepsilon^2)\;
\end{equation}
while
 \begin{equation}
 \big(\textrm{Det}\,D(\varepsilon,t)\big)^{-1/2}
\,=\,1 -\,8\,\mu b^2\varepsilon \int_{\Delta}\hat
g(z,t)\,e^{\left.\varphi_{cl}(t)\right|_{\,\eta\,=\;0}} \,\hat
g(z,z)\,d^2z\,+O(\varepsilon^2)
\end{equation}
which is the quantum determinant contribution occurring in (\ref{2
point b serie }) evaluated on $\eta=0$.\\
Thus, to orders $O(\varepsilon)$ and $O(b^0)$ included, the
logarithm of the  ``normalized'' two point function
(\ref{conformal block}) with $z_1=0$ and $z_2=t$ becomes
\begin{eqnarray}\label{2 point log integrals}
\log\,\frac{\left\langle \, V_{\eta/b}(0)\,
V_{\varepsilon/b}(t)\,\right\rangle}{\rule{0pt}{.41cm}\left\langle
\, V_{\eta/b}(0)\,\right\rangle \left\langle \,
 V_{\varepsilon/b}(t)\,\right\rangle}
 & = & \frac{\varepsilon}{b^2}\;
 \Big\{\,\varphi_{cl}(t)-\left.\varphi_{cl}(t)\right|_{\,\eta\,=\;0}\Big\}\\
 \rule{0pt}{.7cm}& &\hspace{-2.5cm}
- \,8\,\mu b^2\varepsilon \,\left\{\,
\int_{\Delta}\hspace{-.07cm}g(z,t)\,e^{\varphi_{cl}(z)}
g(z,z)\,d^2z\, - \int_{\Delta}\hspace{-.07cm}\hat g(z,t)\,
e^{\left.\varphi_{cl}(t)\right|_{\,\eta\,=\;0}} \,\hat
g(z,z)\,d^2z\,\right\}\,.\nonumber
\end{eqnarray}
The first integral occurring in this expression can be computed by
exploiting the partial wave representation given in appendix
\ref{details green function} (see (\ref{gn series}) and (\ref{gn
factorization})). Because of invariance under rotations, only the
wave $m=0$ contributes and we have
\begin{eqnarray}
- \,8\,\mu b^2 \varepsilon\int_{\Delta}g(z,t)\,e^{\varphi_{cl}(z)}
g(z,z)\,d^2z & = & \\
\rule{0pt}{.9cm}& & \hspace{-7.1cm}- \,8\,\mu b^2 \pi\,
\varepsilon\, \left\{\,b_0(|t|^2)
\int_{0}^{|t|^2}\hspace{-.3cm}a_0(|z|^2)\,e^{\varphi_{cl}(z)}
g(z,z)\,d|z|^2+a_0(|t|^2)
\int_{|t|^2}^{1}\hspace{-.1cm}b_0(|z|^2)\,e^{\varphi_{cl}(z)}
g(z,z)\,d|z|^2\,\right\} \nonumber
\end{eqnarray}
where $g(z,z)$ is given in (\ref{g(z,z)}). These integrals can be
computed explicitly through an integration by parts, while the
other integral occurring in (\ref{2 point log integrals}) is the
tadpole contribution in the geometry
of the pseudosphere without singularities \cite{ZZpseudosphere}.\\
We remark that (\ref{2 point log integrals}) is invariant
under $SU(1,1)$ transformations, as expected. Indeed, since both
classical fields $\varphi_{cl}(t)$ and
$\left.\varphi_{cl}(t)\right|_{\,\eta\,=\;0}$ transform as in
(\ref{varphiB transformations}) under $SU(1,1)$ and $dw/dz$ is
independent of $\eta$, the classical term is invariant. As for the
quantum determinant contribution, by using the transformation law
(\ref{g(z,z) SU(1,1) transformation }) for $g(z,z)$ and $\hat
g(z,z)$, together with the equations for the Green functions
$g(z,t)$ and $\hat g(z,t)$, we find that the variation under
$SU(1,1)$ of the quantum determinant contribution is given by
\begin{equation}
\lim_{r\,\rightarrow\,1}\,\frac{2}{2\pi i}\,\left[\;
\oint_{\partial\Delta_{r}}
\hspace{-.2cm}\partial_z\big(\,g(z,t)-\hat g(z,t)\,\big) \log
\left|\,\frac{dw}{dz}\,\right|^2 \hspace{-.1cm}dz\,+
\oint_{\partial\Delta_{r}} \hspace{-.2cm}\big(\,g(z,t)-\hat
g(z,t)\,\big)\, \log
\frac{d\bar{w}}{d\bar{z}}\;d\bar{z}\,\right]\;.
\end{equation}
Since $g(z,t\hspace{.06cm})-\hat
g(z,t\hspace{.06cm})\,=\,O\big((1-z\bar{z})^2\big)$ when
$|z|\rightarrow 1$, these integrals vanish in the limit
$r\rightarrow 1$; hence the expression (\ref{2 point log
integrals}) is invariant under $SU(1,1)$ transformations and we
can substitute $t\bar{t}$ with the $SU(1,1)$ invariant ratio
$\omega$ in the explicit expression for (\ref{2 point log
integrals}).\\
Thus, including also the classical terms, (\ref{2 point log
integrals}) becomes
\begin{eqnarray}\label{2 point final result}
\log\,\frac{\left\langle \, V_{\eta/b}(z_1)\,
V_{\varepsilon/b}(z_2)\,\right\rangle}{\rule{0pt}{.41cm}\left\langle
\, V_{\eta/b}(z_1)\,\right\rangle \left\langle \,
 V_{\varepsilon/b}(z_2)\,\right\rangle}
 & = &
 \frac{\varepsilon}{b^2}\;
 \left\{-\log\frac{\big(\,\omega^\eta-\omega^{1-\eta}\,\big)^2}
 {\rule{0pt}{.38cm}(1-2\eta)^2}
 \,+\,\log(\hspace{.02cm}1-\omega)^2\,\right\}\nonumber\\
\rule{0pt}{1cm} & & \hspace{-5cm}+\;\varepsilon\,
 \left\{\,\frac{2}{\rule{0pt}{.4cm}\big(1-\omega^{1-2\eta}\big)^2}\,
 \left(\,B_{\omega}(2-2\eta,0)+\psi(2-2\eta)+
\gamma_{\,\scriptscriptstyle\hspace{-.05cm}E} 
         +\frac{1}{2(1-2\eta)}\rule{0pt}{.7cm}\right.\right.
         \\
\rule{0pt}{.7cm} & & \hspace{-1.7cm} +\;\omega^{2(1-2\eta)}
\left(\,B_{\omega}(2\eta,0)+\psi(2\eta)+
\gamma_{\,\scriptscriptstyle\hspace{-.05cm}E}+\frac{3}{2(1-2\eta)}\,
-\log\omega\,\right)       \nonumber\\
\rule{0pt}{.9cm} & &\hspace{-.9cm}\left. \rule{0pt}{.7cm}
\left.+\,2\,\omega^{1-2\eta}\left(\,\log(1-\omega)
-\,\frac{1}{1-2\eta}\;\right)\rule{0pt}{.7cm}\right)
+\,2\,\log(1-\omega)\,-\,3\;\right\}\;.\nonumber
\end{eqnarray}
with $\omega=\omega(z_1,z_2)$ given by (\ref{SU(1,1) invariant
ratio t complex}). According to the analysis performed in
\cite{ZZpseudosphere}, up to the orders $O(b^0)$ and
$O(\varepsilon)$ included, but
 exact in $\eta$ and $\omega(z_1,z_2)$, the formula (\ref{2 point
   final result}) 
gives the expansion of the logarithm of the conformal block
occurring in (\ref{conformal
block}).\\

\noindent The expression (\ref{2 point final result}), which is
exact in $\eta$ and $\omega$ up to orders $O(b^0)$ and
$O(\varepsilon)$ included, provides the summation of two infinite
classes of graphs, ordered according to a power expansion in
$\eta$.\\
In \cite{MTgeometric} the classical part and the quantum
determinant were computed respectively up to
$O(\varepsilon\eta^3/b^2)$ and $O(\varepsilon\eta)$ included, by
the explicit computation of every single graph. The procedure
presented here extends largely the results obtained in
\cite{MTgeometric} because it allows to get directly the sum of
infinite classes of graphs, from which one can find the
contribution of all the graphs occurring at a given perturbative
order without computing them separately.\\
 The classical part at $O(\varepsilon)$ gives
\begin{eqnarray}
\frac{\varepsilon}{b^2}\;
 \left\{-\log\frac{\big(\,\omega^\eta-
 \omega^{1-\eta}\,\big)^2}{\rule{0pt}{.38cm}(1-2\eta)^2}
 \,+\,\log(1-\omega)^2\,\right\}
& = &\,\frac{\varepsilon\,\eta}{b^2}\; \left\{\rule{0pt}{.4cm}\;
\begin{minipage}[c]{1.6cm}
  \includegraphics[width=1.6cm]{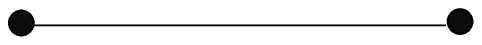}
\end{minipage}\hspace{0cm}\;\right\}\,
\\
\rule{0pt}{1cm} & & \hspace{-3.2cm} +\,
\frac{\varepsilon\,\eta^2}{b^2}\, \left\{\rule{0pt}{.6cm}\;
\begin{minipage}[c]{1.6cm}
  \includegraphics[width=1.6cm]{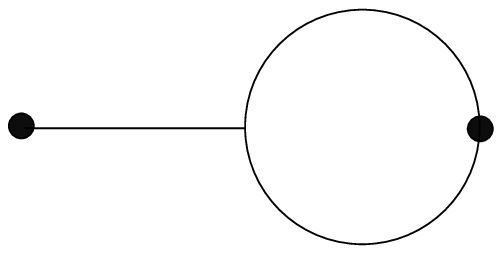}
\end{minipage}\hspace{0cm}\;\right\}\,+\,
\frac{\varepsilon\,\eta^3}{b^2}\, \left\{\rule{0pt}{.6cm}\;
\begin{minipage}[c]{4cm}
  \includegraphics[width=4cm]{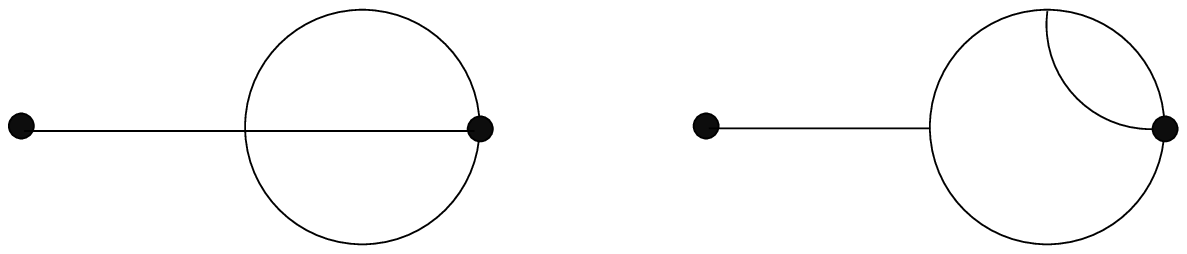}
\end{minipage}\hspace{0cm}\;\right\}
\nonumber\\
\rule{0pt}{1.7cm}
 & & \hspace{-3.2cm}+\,\frac{\varepsilon\,\eta^4}{b^2}\,
\left\{\rule{0pt}{1.2cm}\;
\begin{minipage}[c]{6cm}
  \includegraphics[width=6cm]{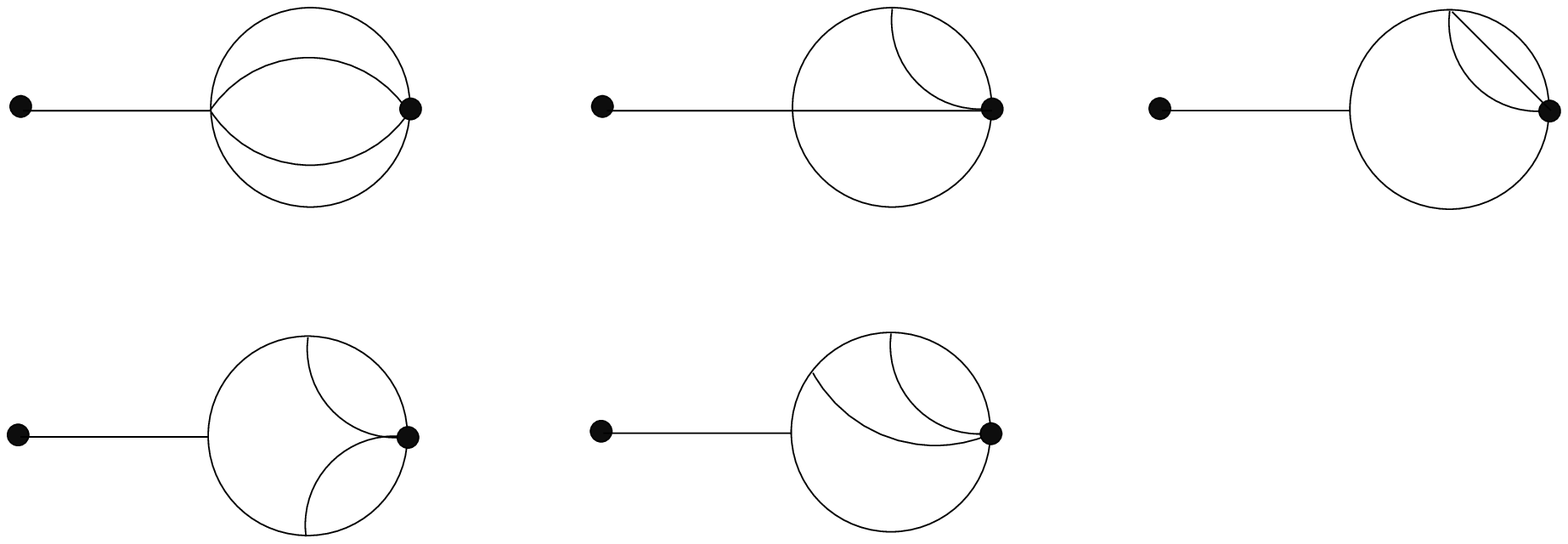}
\end{minipage}\hspace{0cm}\;\right\}\,+\,\dots
\nonumber\\
\rule{0pt}{1.3cm} & & \hspace{-7.1cm} =\;4\;
\frac{\varepsilon\,\eta}{b^2}\;\hat g(z_1,z_2) +\,4\;
\frac{\varepsilon\,\eta^2}{b^2}\left(\,\frac{2\,\omega\,\log\omega
}{\rule{0pt}{.4cm}(1-\omega)^2}\,-1\right) -\,\frac{8}{3}\;
\frac{\varepsilon\,\eta^3}{b^2}\left(\,\frac{3\,\omega\,(1+\omega)
\,\log\omega }{\rule{0pt}{.4cm}(1-\omega)^3}\,+2\right)
\nonumber\\
\rule{0pt}{1cm} & & \hspace{-6.5cm} +\,\frac{4}{3}\;
\frac{\varepsilon\,\eta^4}{b^2}\left(\,\frac{4\,\omega\,
(1+4\,\omega+\omega^2) \,\log\omega
}{\rule{0pt}{.4cm}(1-\omega)^4}\,-6\right)+\,\dots
\end{eqnarray}
while the quantum determinant contribution at $O(\varepsilon)$
provides the following infinite class of perturbative graphs
\begin{eqnarray}\label{2 point quantum det}
\rule{0pt}{1cm} & & \hspace{-1cm}\varepsilon\,
 \left\{\,\frac{2}{\rule{0pt}{.4cm}\big(1-\omega^{1-2\eta}\big)^2}\,
 \left(\,B_{\omega}(2-2\eta,0)+\psi(2-2\eta)+
\gamma_{\,\scriptscriptstyle\hspace{-.05cm}E}
         +\frac{1}{2(1-2\eta)}\rule{0pt}{.7cm}\right.\right.
         \\
\rule{0pt}{.7cm} & & \hspace{1.3cm} +\,\omega^{2(1-2\eta)}
\left(\,B_{\omega}(2\eta,0)+\psi(2\eta)+
\gamma_{\,\scriptscriptstyle\hspace{-.05cm}E}+\frac{3}{2(1-2\eta)}\,
-\log\omega\,\right)       \nonumber\\
\rule{0pt}{.9cm} & &\hspace{2.3cm}\left. \rule{0pt}{.7cm}
\left.+\,2\,\omega^{1-2\eta}\left(\,\log(1-\omega)
-\,\frac{1}{1-2\eta}\;\right)\rule{0pt}{.7cm}\right)
+\,2\,\log(1-\omega)\,-\,3\;\right\}\;\;=\nonumber\\
\rule{0pt}{1cm} & & \hspace{-1cm} =\;\;\varepsilon\,\eta\,
\left\{\rule{0pt}{.6cm}\;
\begin{minipage}[c]{6.5cm}
  \includegraphics[width=6.5cm]{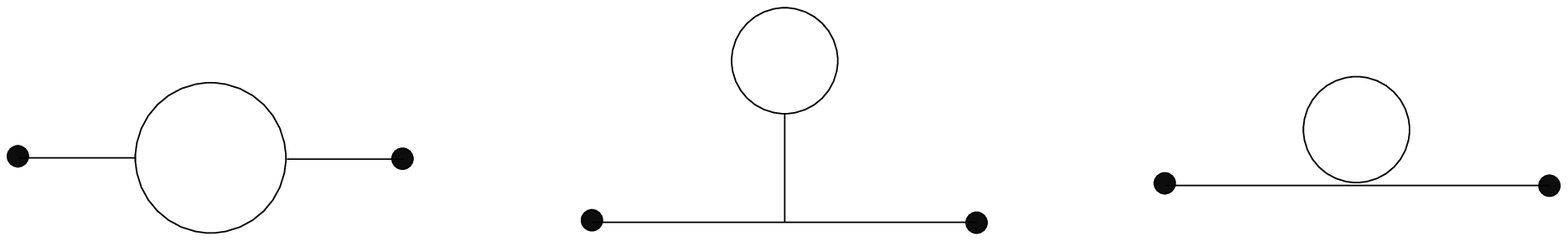}
\end{minipage}\hspace{0cm}\;\right\}
\nonumber\\
\rule{0pt}{2cm}
 & & \hspace{-1cm}+\;\varepsilon\,\eta^2\,
\left\{\rule{0pt}{1.2cm}\;
\begin{minipage}[c]{14cm}
  \includegraphics[width=14cm]{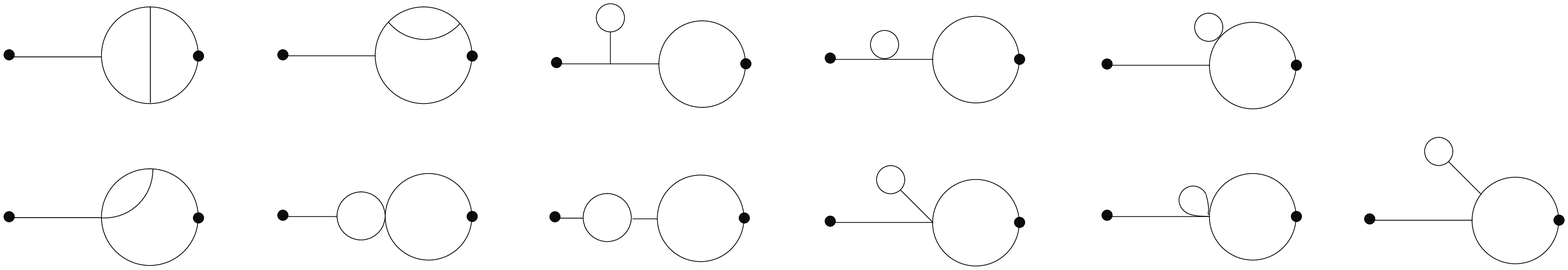}
\end{minipage}\hspace{0cm}\;\right\}
\nonumber\\
\rule{0pt}{.7cm}
 & & \hspace{-1cm}+\,\dots\nonumber\\
\rule{0pt}{1cm} & &\hspace{-1cm}
=\;\;2\,\varepsilon\,\eta\,\left(\,3\,+\,\frac{2\,\omega^2
\log\omega}{(1-\omega)^2}\,-\,
2\;\frac{1+\omega}{1-\omega}\;\textrm{Li}_2(1-\omega)\right)
\nonumber\\
\rule{0pt}{1cm} & & \hspace{-.4cm}
+\;\frac{4\,\varepsilon\,\eta^2}{(1-\omega)^2}\,
\left(\,3\,-2\,\big(1+\omega^2\big)\,\zeta(3)\,+\,\frac{2}{3}\,\pi^2
\,\omega\,\log\omega\,-\,\frac{\omega^2\, (5+\omega)\, \log\omega
}{(1-\omega)}\,\right.\nonumber\\
\rule{0pt}{.8cm} & & \hspace{-2.5cm}\left.\phantom{\frac{\omega^2
\log(\hspace{.02cm}t\bar{t}\hspace{.05cm})^3
}{\rule{0pt}{.4cm}3\,(\hspace{.02cm}1-t\bar{t}\hspace{.04cm})}}
-\,2\,\big(1+4\,\omega+\omega^2\big)\,\log(1-\omega)\,
\log\omega\,-\,2\,(1+\omega)^2\log\omega\,
\textrm{Li}_2(\omega)\,+\, 2\,\big(1+\omega^2\big)\,
\textrm{Li}_3(\omega)\,\right)\nonumber\\
\rule{0pt}{.5cm} & & \hspace{-.4cm}+\,\dots\nonumber
\end{eqnarray}
where $\textrm{Li}_\nu(x)$ is the polylogarithm function.\\

\noindent Notice that, by using (\ref{F21 psi serie}), the
behavior of (\ref{2 point final result}) when $\omega(z_1,z_2)
\rightarrow 1$ is
\begin{equation}\label{cluster behavior}
    \log\,\frac{\left\langle \, V_{\eta/b}(z_1)\,
V_{\varepsilon/b}(z_2)\,\right\rangle}{\rule{0pt}{.41cm}\left\langle
\, V_{\eta/b}(z_1)\,\right\rangle \left\langle \,
 V_{\varepsilon/b}(z_2)\,\right\rangle}\;=\;
 \left(\,\frac{\eta\,(1-\eta)}{3}\;\frac{\varepsilon}{b^2}\,-\,
\frac{\eta\,(1-7\eta)}{18}\;\varepsilon\,\right)
 (1-\omega)^2+O\big((1-\omega)^3\big)
\end{equation}
i.e. $g_{\,\eta/b,\,\varepsilon/b}(\omega)\rightarrow 1$ for the
normalized two point function (\ref{conformal block}).\\
The fact that $\left\langle  V_{\eta/b}(z_1)\,
V_{\varepsilon/b}(z_2)\right\rangle \rightarrow \left\langle
V_{\eta/b}(z_1)\right\rangle \left\langle
V_{\varepsilon/b}(z_2)\right\rangle$ when $\omega(z_1,z_2)
\rightarrow 1$, i.e. when the geodesic distance diverges, is the
cluster property and it is the boundary condition used in the
bootstrap approach \cite{ZZpseudosphere} to characterize the
pseudosphere.\\

\noindent As a further check of our result (\ref{2 point final
result}), we consider the auxiliary bulk two point function
$\left\langle
V_{-1/(2b)}(z_1)\,V_{\varepsilon/b}(z_2)\right\rangle$ containing
the primary field $V_{-1/(2b)}(z_1)$, which is degenerate at level
2. When the two point function contains a degenerate primary field
at level 2, it satisfies a second order linear differential
equation and it can be determined explicitly \cite{difrancesco}.
In our case, we have that \cite{FZZ}
\begin{equation}
g_{\,-1/(2b),\,\varepsilon/b}(\omega)\,=\,
\omega^{\varepsilon/b^2}\, _2
F_1\big(\,1+1/b^2,\,2\varepsilon/b^2;\,2+2/b^2;\,1-\omega\,\big)
\end{equation}
Expanding the logarithm of this expression, we find
\begin{equation}\label{expansion degenerate}
    \log\big[\,g_{\,-1/(2b),\,\varepsilon/b}(\omega)\,\big]\,=\,
    \frac{\varepsilon}{b^2}\,\big(\,\log\omega\,+\,2\,\log 2\,-\,
    2\,\log(1+\omega)\,\big)
    +\,\frac{\varepsilon}{2}\,\left(\,\frac{1-\omega}{1+\omega}\,\right)^2+\,
    O(\varepsilon b^2)
\end{equation}
up to $O(\varepsilon)$ and $O(b^0)$ included. The expansion
(\ref{expansion degenerate}) agrees with our expansion (\ref{2
point final result}) evaluated for $\eta=-1/2$.

\section*{Conclusions}

\noindent We have obtained the one and two point functions on the
pseudosphere with heavy charges to one loop. For the one point
function agreement is found with the bootstrap formula given by ZZ
\cite{ZZpseudosphere} while the two point function provides a new
expression for the case of one
finite charge and an infinitesimal one.\\
Furthermore we have proved that the correct quantum dimensions
recovered to one loop are left unchanged to all orders perturbation theory.\\
In the next publication \cite{MTboundary} we extend the present
approach to the conformal boundary case.

\section*{Appendices}

\appendix
\section{The background field at infinity}
\label{appendix phiB}

In this appendix we examine the behavior of the classical
background on the pseudosphere at infinity in presence of $N$
sources. We already saw in section \ref{classical} that the one
source solution (\ref{phiclassic}) behaves at infinity like
\begin{equation}
\varphi_{cl}(z) \,=\, -\log(1-z\bar z)^2 + {\rm const} +
O\big((1-z\bar z)^2\big)
\end{equation}
i.e. no $O(1-z\bar z)$ term occurs. This is relevant to have the
second line of (\ref{quantumaction}) vanishing for a quantum field
$\chi$ which behaves like $\log(1-z \bar z)$ when $|z|\rightarrow
1$. Here we prove that the term $O(1-z\bar z)$ is absent also in
the background $\varphi_{\scriptscriptstyle\hspace{-.05cm}B}$
generated by $N$ sources. Since it is simpler to work in the
$\mathbb{H}$ representation, we begin by using it.\\
Being $\widetilde{Q}(\xi)$ a real function, we can choose two real
independent solutions $\tilde y_j(\xi)$ of
\begin{equation}
\tilde y_j''(\xi) + \widetilde{Q}(\xi) \tilde
y_j(\xi)=0\hspace{2cm}j\,=\,1,2\;.
\end{equation}
Using the fact the the wronskian is different from zero, it is
simple to prove the following lemma: the identical vanishing of $a
\tilde y_1^2+b\tilde y_2^2+c \tilde y_1 \tilde y_2$ implies
$a=b=c=0$. As explained in section \ref{classical}, the solution
of the Liouville equation is given by
\begin{equation}
\pi\mu b^2
e^{\tilde{\varphi}_{\scriptscriptstyle\hspace{-.02cm}B}(\xi)}
=\frac{|w_{12}|^2}{F^2}
\end{equation}
where the most general form for $F$ is
\begin{equation}
F\,=\,\big(\alpha\, \tilde{y}_1(\xi)+\beta\,
\tilde{y}_2(\xi)\big)\big(\bar\alpha
\,\tilde{y}_1(\bar\xi)+\bar\beta \,\tilde{y}_2(\bar\xi)\big)\, -\,
\big(\gamma \,\tilde{y}_1(\xi)+\delta
\,\tilde{y}_2(\xi)\big)\big(\bar\gamma\,
\tilde{y}_1(\bar\xi)+\bar\delta \,\tilde{y}_2(\bar\xi)\big)
\end{equation}
The identical vanishing of $F$ for real $\xi$ implies, as a
consequence of the previous lemma, $\gamma=\bar\alpha$ and
$\delta=\bar\beta$, hence we can write
\begin{equation}
F=Y(\xi)
\overline{\rule{0pt}{.33cm}Y}(\bar{\xi})-\overline{\rule{0pt}{.33cm}Y}(\xi)
Y(\bar{\xi})\;.
\end{equation}
We notice that $F$ is odd in $\textrm{Im}\xi$, so that for small
$\textrm{Im}\xi$ we have
\begin{equation}\label{behavior}
\frac{1}{F^2} \,=\,
\frac{1}{4\big(\,\textrm{Im}\xi+O\big((\textrm{Im}\xi)^3\big)\big)^2}
\,=\,\frac{1}{4(\textrm{Im}\xi)^2}\,+\,O(1)\,.
\end{equation}
On the other hand, if we start from the $\Delta$ representation,
being $w_{12}$ invariant, for $z\bar{z}\rightarrow 1$ we have
\begin{eqnarray}\label{infinity delta}
\pi\mu b^2 e^{\varphi_{\scriptscriptstyle\hspace{-.02cm}B}(z)} &=&
\frac{|w_{12}|^2}{\left[\,(1-z\bar z)+c_1 (1-z\bar
z)^2+O\big((1-z\bar{z})^3\big)\,\right]^2}\nonumber\\
\rule{0pt}{.9cm} &=&|w_{12}|^2\left(\frac{1}{(1-z\bar
z)^2}\,-\,\frac{2c_1}{(1-z\bar z)}\,+\,O(1)\right)\;.
\end{eqnarray}
Taking into account of the jacobian, (\ref{infinity delta}) gives
\begin{equation}
\frac{1}{F^2}
\,=\,\frac{4}{|\xi+i|^4}\,\left(\,\frac{|\xi+i|^4}{16
(\textrm{Im}\xi)^2}-\frac{c_1|\xi+i|^2}{4
\,\textrm{Im}\xi}\,+\,O(1)\right)\,=\,
\frac{1}{4(\textrm{Im}\xi)^2}-\frac{c_1}{|\xi+i|^2\,\textrm{Im}\xi
}\,+ \,O(1)
\end{equation}
which gives $c_1=0$, when compared to (\ref{behavior}).

\section{Details about the Green function}
\label{details green function}

\noindent In this appendix we outline some technical details
necessary to compute the explicit form of the Green function.\\
By exploiting the $SU(1,1)$ invariance, we can set the source with
charge $\eta/b$ in the origin and the
one with charge $\varepsilon/b$ in $t \in (-1,1)\setminus{0}$. \\
By using the definition (\ref{Iij(z)}) and the expression of
$q(z)$ given in (\ref{q on Delta}), we can perform explicitly the
integrals $I_{ij}(z)$ in the $\Delta$ representation of the
pseudosphere. They are given by
\begin{eqnarray}
I_{12}(z)\hspace{-.1cm} & = & \hspace{-.1cm}-\,\frac{t
z+z/t-2}{\rule{0pt}{.4cm}(z-t)\,(z-1/t)}\,+\,
C(\eta,t,\beta\hspace{.03cm})\,\Big(
\log(\,z-t\,)-\log(\,z-1/t\,)-\log t^2\,\Big)-\,2
\nonumber\\
& & \\
\rule{0pt}{1cm} I_{11}(z) \hspace{-.1cm}& = &\hspace{-.1cm}
\,z^{2\eta-1}\,\left\{\,-\,z\,\frac{2z-t-1/t}{\rule{0pt}{.4cm}(z-t)\,(z-1/t)}\,
-\,\frac{A(\eta,t,\beta)}{2\eta}\;\frac{z}{t}\,
F(\,2\eta,1;1+2\eta\,;z/t\,)\,\right.\nonumber \\
& & \hspace{5.2cm}
\phantom{\Bigg\{}+\,\frac{B(\eta,t,\beta)}{2\eta}\;z\,t\,
F(\,2\eta,1;1+2\eta\,;z\,t\,)\,\Bigg\}
\\
& & \nonumber \\
\rule{0pt}{1cm} I_{22}(z) \hspace{-.1cm}& = &\hspace{-.1cm}
\,z^{1-2\eta}\,\left\{\,-\,z\,\frac{2z-t-1/t}{\rule{0pt}{.4cm}(z-t)\,(z-1/t)}\,
-\,\frac{A(1-\eta,t,\beta)}{2(1-\eta)}\;\frac{z}{t}\,
F(\,2-2\eta,1;3-2\eta\,;z/t\,)\,\right.\nonumber \\
& & \hspace{5.2cm}
\phantom{\Bigg\{}+\,\frac{B(1-\eta,t,\beta)}{2(1-\eta)}\,z\,t\,
F(\,2-2\eta,1;3-2\eta\,;z\,t\,)\,\Bigg\}
\nonumber\\
& &
\end{eqnarray}
where the functions $A(\eta,t)$, $B(\eta,t)$ and $C(\eta,t)$ are
\begin{eqnarray}
A(\eta,t,\beta) & = &
2\,\frac{\eta+(1-\eta)\,t^2}{1-t^2}\,+\,\frac{t\,\beta}{1-t^2}\;\;=
\;\;B(1-\eta,t,\beta)\\
\label{C(eta,t)} \rule{0pt}{.9cm} C(\eta,t,\beta) & = &
\frac{1+t^2+\beta\,t}{1-t^2}\;\;=\;\;\frac{A(\eta,t,\beta)+
A(1-\eta,t,\beta)}{2}\;.
\end{eqnarray}
Notice that only the hypergeometric function of type
$F(a,1;a+1\,;x)$ occurs. It is related to a special case of the
incomplete beta function \cite{Bateman,Prudnikov}
\begin{equation}
    \frac{x^a}{a}\,F(a,1;a+1\,;x)\,=\,B_{x}(a,0)\,=\,\int_0^{\,x}
    \frac{y^{a-1}}{1-y}\;dy\;.
\end{equation}
Now, by inserting the expressions of $I_{ij}(z)$ into
(\ref{g(z,t)}), we find the explicit expression for the propagator
\begin{eqnarray}\label{g(z,t) A,B,C}
g(z,t) & = & -\,\frac{1}{2\,(1-2\eta)}\;
\frac{1}{(z\bar{z})^{\eta}-(z\bar{z})^{1-\eta}}\;\times\\
& & \rule{0pt}{.9cm}
 \hspace{-.8cm} \times\;\left\{\;
\Big((z\bar{z})^{\eta}+(z\bar{z})^{1-\eta}\Big)\,
\left[\;\rule{0pt}{.58cm}C(\eta,t,\beta)\,\Big(\log \omega(z,t)-
\log t^2\Big) + 4 + 2 \,\textrm{Re}\,h(t)\;\right]\right.
\nonumber \\
\rule{0pt}{.7cm}
 & & \hspace{.4cm}
 +\,\frac{(z\bar{z})^{\eta}}{2\eta}\;
 \left[\;A(\eta,t,\beta)\,\Big(z/t\,F(2\eta,1;1+2\eta;z/t)+\textrm{c.c.}\Big)
        \rule{0pt}{.58cm}\right. \nonumber\\
 \rule{0pt}{.7cm}& & \hspace{5cm}\left.\rule{0pt}{.58cm}
        -B(\eta,t,\beta)\,\Big(z\,t\,
F(2\eta,1;1+2\eta;z\,t)+\textrm{c.c.}\Big)\;\right]
\nonumber \\
\rule{0pt}{.7cm}
 & & \hspace{.4cm}
 +\,\frac{(z\bar{z})^{1-\eta}}{2(1-\eta)}\;
 \left[\;A(1-\eta,t,\beta)\,\Big(z/t\,
F(2-2\eta,1;3-2\eta;z/t)+\textrm{c.c.}\Big)
       \rule{0pt}{.58cm}\right.
\nonumber \\
\rule{0pt}{.7cm} & & \hspace{3cm}
\left.\left.\rule{0pt}{.58cm}\phantom{\frac{X}{X}}
 -\,B(1-\eta,t,\beta)\,\Big(z\,t\,
F(2-2\eta,1;3-2\eta;z\,t)+\textrm{c.c.}\Big)\;\right]\;\right\}
\nonumber
\end{eqnarray}
where $\omega(z,t)$ is the $SU(1,1)$ invariant
\begin{equation}\label{SU(1,1) invariant ratio}
\omega(z,t)\,=\,\left|\,\frac{z-t}{1- z\,t}\,\right|^2\hspace{2cm}
t\,\in\,(-1,1)
\end{equation}
which is related to the geodesic distance on the pseudosphere
without sources.\\
We remark that the expression (\ref{g(z,t) A,B,C}) satisfies the
equation
\begin{equation}\label{g(z,t) equation}
    -\,\frac{2}{\pi}\,\partial_z\partial_{\bar{z}}\,g(z,t)\,+ \,4\,\mu
    b^2\,e^{\varphi_{cl}}\,g(z,t) \,=\, \delta^2(z-t)
\end{equation}
for any $\beta$. By employing the expansion \cite{Prudnikov}
\begin{equation}\label{F21 psi serie}
F(\,a,1;a+1;\,w\,)
\,=\,a\sum_{k\,=\,0}^{\infty}\frac{(a)_{k}}{k!}\,
\Big(\psi(1+k)-\psi(a+k)-\log(1-w)\Big)\,(1-w)^k
\end{equation}
where $(a)_{k}=a(a+1)\dots(a+k-1)=\Gamma(a+k)/\Gamma(a)$ is the
Pochhammer symbol and $\psi(x)=\Gamma'(x)/\Gamma(x)$, one can see
that for any $\beta$ the logarithmic divergence of $g(z,t)$ when
$z \rightarrow t$ is exactly $-1/2 \,\log|z-t|^2$ because the
following identity
\begin{equation}\label{deltadivergence}
\frac{\big(1+t^{2\,(1-2\eta)}\big)\,C(\eta,t,\beta)
-A(\eta,t,\beta)-A(1-\eta,t,\beta)\,t^{2(1-2\eta)}}{
\rule{0pt}{.4cm}(1-2\eta)\big(1-t^{2\,(1-2\eta)}\big)}\,=\,1\;.
\end{equation}
To determine $\beta$, we impose the monodromy of $g(z,t)$ around
$z = t$. When $t<|z|<1$, by using the identity \cite{Bateman}
\begin{equation}\label{F21 1/w}
    F(\,a,1;a+1;\,1/w\,) \,=\, \frac{a\pi}{\sin (a\pi)}\,(-\,w)^a
    +\frac{a}{1-a}\;w\,F(\,1-a,1;2-a;\,w\,)
\end{equation}
we find that $(-w)^a$ introduces a term that breaks the monodromy.
The vanishing of this term leads to the equation
\begin{equation}\label{beta eq. monodromy}
A(\eta,t,\beta)-t^{2(1-2\eta)} A(1-\eta,t,\beta)\,=\,0
\end{equation}
which allows to get
\begin{equation}\label{beta appendix}
    \beta \,=\,
-\,2\; \frac{\eta+(\,1-\eta\,) \,t^2-t^{2\,(1-2\eta)}
\left(\,1-\eta+\eta\,t^2\,\right)}{\rule{0pt}{.4cm}t\,
\big(\,1-t^{2\,(1-2\eta)}\big)}
\end{equation}
which solves the problem of finding the $O(\varepsilon)$ terms
of the accessory parameters in the perturbed geometry.\\
The expression for $\textrm{Re}\,h(t)$ can be obtained by studying
the asymptotic behavior of the Green function $g(z,t)$ when $|z|
\rightarrow 1$. To get this result, we set
$z\hspace{-.1cm}=\hspace{-.1cm}e^{i\theta}$ in (\ref{g(z,t)
A,B,C}) and we analyze its leading term. By using the identity
(\ref{F21 1/w}) and the following one \cite{Bateman}
\begin{equation}
    F(\,a,1;a+1;\,w\,) \,=\, \frac{a}{a-1}\;\frac{1}{w}\;
    \Big(\,F(\,a-1,1;a;\,w\,)-1\,\Big)
\end{equation}
we get
\begin{equation}\label{F21 1/w evoluta}
    F(\,a,1;a+1\,;1/w\,) \,=\, a\,\left(\,\frac{\pi}{\sin (a\pi)}\,(-w)^a+
    \frac{w}{1-a}+\frac{w^2}{2-a}\,F(\,2-a,1;3-a\,;w\,)\right)
\end{equation}
which allows us to reduce all the hypergeometric functions
occurring in (\ref{g(z,t) A,B,C}) to hypergeometric functions with
the same parameters but different variables. Then, for any $\beta$
we find that the leading order in $(1-z\bar{z})$ of $g(z,t)$
contains no contributions from the hypergeometric functions but it
includes a term $(-w)^a$ which would break the monodromy. The
coefficient of such a term vanishes because of the explicit
expression for $\beta$ found before.\\
Thus, we have that the leading order in $(1-z\bar{z})$ of the
expression contained between the curly brackets in (\ref{g(z,t)
A,B,C}) is
\begin{equation}\label{Reh(t) equation}
    2\,\Big(\,2\,\textrm{Re}\,h(t)-C(\eta,t,\beta\hspace{.03cm})
\log t^2-2 \,\Big)
\end{equation}
where $\beta$ is given by (\ref{beta appendix}). By requiring the
vanishing of (\ref{Reh(t) equation}), we find the explicit
expression of $\textrm{Re}\,h(t)$
\begin{equation}\label{Reh(t) appendix}
\textrm{Re}\,h(t) \,=\,
\frac{1}{2}\,\left(\,\frac{1+t^{2\,(1-2\eta)}}{\rule{0pt}{.4cm}
1-t^{2\,(1-2\eta)}}\,\log t^{2\,(1-2\eta)}+2\,\right)
\end{equation}
which is also given in (\ref{Reh(t)}). The vanishing of
(\ref{Reh(t) equation}) is necessarily imposed by the fact that
the divergence of the field $\varphi$ when $|z| \rightarrow 1$ is
at most logarithmic.\\
We will find later that $g(z,t)$ vanishes quadratically when $|z|
\rightarrow 1$. One could see this also at this level, but it is
much simpler once the expansion of $g(z,t)$
in partial waves is available.\\
By exploiting the invariance in value of the Green function under
rotations, we can easily generalize our formula to the case of
complex $t\in\Delta$. Thus, we have that $t^2$, $z/t$, $\bar{z}/t$
$z\,t$, and $\bar{z}\,t$ become respectively $t\bar{t}$, $z/t$,
$\bar{z}/\,\bar{t}$,
$z\,\bar{t}$, and $\bar{z}\,t$.\\
The final expression for the Green function is
\begin{eqnarray}\label{g(z,t) final form}
 g(z,t) & = & -\,\frac{1}{\rule{0pt}{.4cm}2}\;
\frac{1+(z\bar{z})^{1-2\eta}}{\rule{0pt}{.4cm}1-(z\bar{z})^{1-2\eta}}
\;\left\{\;\frac{1+(t\bar{t}\hspace{.04cm})^{1-2\eta}}
{\rule{0pt}{.4cm}1-(t\bar{t}\hspace{.04cm})^{1-2\eta}}
\;\log\omega(z,t) +\,\frac{2}{\rule{0pt}{.4cm}1-2\eta}\;\right\}
\\
\rule{0pt}{.9cm} & & -\,
\frac{1}{\rule{0pt}{.4cm}\,1-(z\bar{z})^{1-2\eta}}\;
\frac{1}{\rule{0pt}{.4cm}\,1-(t\bar{t}\hspace{.04cm})^{1-2\eta}}\,\times
\nonumber \\
\rule{0pt}{.9cm} & & \hspace{-2cm}\times \;
\left\{\;\frac{(t\bar{t}\hspace{.04cm})^{1-2\eta}}{2\eta}\;
\frac{z}{t}\,F(\,2\eta,1;1+2\eta
;\,z/t\,)+\,\frac{(z\bar{z})^{1-2\eta}}{2(1-\eta)}\;
\frac{z}{t}\,F(\,2-2\eta,1;3-2\eta;\,z/t\,)\,+\,\textrm{c.c.}\;\right.
\nonumber \\
\rule{0pt}{1cm} & &\hspace{-2cm}\left.
-\,\frac{1}{2\eta}\;z\,\bar{t}\,F(\,2\eta,1;1+2\eta;\,z\,\bar{t}\,)\,-\,
\frac{(z\bar{z})^{1-2\eta}(t\bar{t}\hspace{.04cm})^{1-2\eta}}{2(1-\eta)}
\;z\,\bar{t}\,F(\,2-2\eta,1;3-2\eta;\,z\,\bar{t}\,)\,+
\,\textrm{c.c.}\,\right\}\,.
\nonumber
\end{eqnarray}
By using (\ref{F21 1/w evoluta}), this Green function can be
written also in the explicit symmetric form given in (\ref{g(z,t)
symmetric}). Notice that $g(z,t)$ is regular at $z=0$, as we
expect.\\
It is easy to see that $g(z,t)$ is invariant under $\eta
\rightarrow 1-\eta$, which is the semiclassical version of the
duality $\alpha\rightarrow Q-\alpha$. Here it is only a formal
invariance because, due to the finite area condition around the
sources, $\eta < 1/2$ must hold.\\

\noindent A particular case of the Green function (\ref{g(z,t)
final form}) is given by the limit $\eta \rightarrow 0$, which
recovers the geometry of the pseudosphere without curvature
singularities. To compute this limit, we use
\begin{equation}
F(\,a,1;a+1;\,w\,) \,=\, 1 - a\,\log
(\hspace{.03cm}1-w\hspace{0cm}) + \sum_{k\,\geqslant \,2}
(-1)^{k+1} a^k \,\textrm{Li}_k(w)
\end{equation}
and
\begin{equation}
F(\,2-a,1;3-a;\,w\,) \,=\, -\,\frac{2}{w^2}\,\Big(\log (1-w) +
w\Big)+\frac{1}{w^2}\sum_{k\,\geqslant \,1}
a^k\Big(2\,\textrm{Li}_{k+1}(w)-\textrm{Li}_{k}(w)-w\Big)\;.
\end{equation}
After some algebraic manipulation, we find the following $SU(1,1)$
invariant expression
\begin{equation}\label{ZZ propagator}
    \lim_{\eta \,\rightarrow \,0}
    \,g(z,t)\,=\,-\,\frac{1}{2}\,
    \left(\,\frac{1+\,\omega}{1-\,\omega}\, \log \,\omega \,
    +2\,\right)\,\equiv\,\hat g(z,t)\hspace{.6cm},
    \hspace{.6cm}
    \omega\,=\,\left|\,\frac{z-t}{\rule{0pt}{.4cm}1-z\,\bar{t}}\,\right|^2\;.
\end{equation}
This is the propagator on the regular pseudosphere \cite{DFJ,
ZZpseudosphere}, whose classical background is
\begin{equation}\label{phiclassic eta=0}
    e^{\left.\varphi_{cl}\right|_{\eta\,=\,0}}\,=\,
    \frac{1}{\pi\mu b^2(1-z\bar{z})^2}\;.
\end{equation}

\noindent To close this appendix, we provide the partial wave
expansion of the propagator $g(z,t)$, given in (\ref{g(z,t)
symmetric}) or (\ref{g(z,t) final form}). For $x \in \mathbb{R}$,
we have that
\begin{equation}\label{expansion log}
    \log \big(\,1-2x \cos\theta +x^2\,\big)\,=\,
    -\,2\sum_{m\,=\,1}^{\infty}
    \,\frac{x^m}{m}\,\cos(m\theta)
    \hspace{1.2cm}
    x^2 \leqslant 1\;,\hspace{.6cm}
    x\,\cos\theta \neq 1
\end{equation}
and
\begin{equation}\label{expansion 2F1}
\frac{xe^{i\theta}}{a}\,
F(\,a,1;a+1;\,xe^{i\theta}\,)\,+\,\textrm{c.c.}\,=\,
\,2\sum_{m\,=\,1}^{\infty} \,\frac{x^m}{a-1+m}\,\cos(m\theta)
    \hspace{1.2cm}
    0 \leqslant x \leqslant 1\;.
\end{equation}
By using these expansions, we can write the propagator
(\ref{g(z,t) final form}) as a Fourier series as follows
\begin{equation}\label{gn series}
    g(z,t)\,=\,\sum_{m\,=\,0}^{\infty} g_m(x,y)\,\cos(m\theta)
    \hspace{1.7cm} x\,=\,|z|^2\;, \;\;\; y\,=\,|t|^2\;
\end{equation}
where $\theta= \textrm{arg}(z)-\textrm{arg}(t)$. The Fourier
coefficients can be written in the symmetric and factorized form
\begin{equation}\label{gn factorization}
   g_m(x,y)\,=\,\theta(y-x)\,a_m(x)\,b_m(y) +
   \theta(x-y)\,a_m(y)\,b_m(x)
\end{equation}
where the wave with $m=0$ is given by
\begin{equation}
\rule{0pt}{.7cm} a_0(x) \,=\, \frac{1+x^{1-2\eta}}{1-x^{1-2\eta}}
\hspace{1.4cm} b_0(y) \,=\, -\,\frac{1}{2(1-2\eta)}\,
\left(\,\frac{1+y^{1-2\eta}}{1-y^{1-2\eta}}\,\log y^{1-2\eta} +
2\,\right)
\end{equation}
while $a_m(x)$ and $b_m(y)$ for $m\geqslant 1$ read
\begin{eqnarray}
\rule{0pt}{.6cm} a_m(x) \hspace{-.1cm}& = &\hspace{-.1cm}
\frac{x^{m/2}}{1-x^{1-2\eta}}\,\left(\,1-\frac{m-(1-2\eta)}{m+(1-2\eta)}\;
x^{1-2\eta}\,\right) \\
\rule{0pt}{1cm} b_m(y)\hspace{-.1cm} & = &\hspace{-.1cm}
-\,\frac{y^{-m/2}}{m\big(m-(1-2\eta)\big)}\,\left(\,(1-2\eta)\,
\frac{1+y^{1-2\eta}}{1-y^{1-2\eta}}
\,(1-y^m) - m (1+y^m)\,\right)\;.\hspace{1cm}
\end{eqnarray}
This expansion allows to find the asymptotic behavior of the Green
function (\ref{g(z,t) final form}) at infinity in a simple way.
Indeed, since for any $m$
\begin{equation}
b_m(y)\,=\,O\big((1-y)^2\big)
\hspace{.9cm}\textrm{when}\hspace{.8cm} y\rightarrow 1
\end{equation}
then also the propagator vanishes quadratically at infinity
\begin{equation}
g(z,t\hspace{.06cm})\,=\,O\big((1-z\bar{z})^2\big)
\hspace{.9cm}\textrm{when}\hspace{.8cm} |z|\rightarrow 1\;.
\end{equation}
From the behavior
\begin{equation}
a_m(y)\,\propto\,\frac{1}{1-y}
\hspace{.9cm}\textrm{when}\hspace{.8cm} y\rightarrow 1
\end{equation}
we see that $g(z,t)$ given in (\ref{g(z,t) final form}) is the
unique Green function which does not diverge at infinity.\\
The Fourier expansion simplifies also the analysis of the limit
$\eta \rightarrow 0$. Indeed, taking the expressions of $a_m(x)$
and $b_m(y)$ in this limit, with a special care for the case
$m=1$, one can easily verify that they reproduce the Fourier
expansion of the propagator (\ref{ZZ propagator}), which was found
in \cite{MTtetrahedron} and was used there to perform a three loop
calculation on the background of the regular pseudosphere.

\end{document}